\documentclass[useAMS,usenatbib]{mnras}
\usepackage{graphicx,psfig,cite,epsf}  
\usepackage{times}
\usepackage{subfigure}
\usepackage{textcomp}
\usepackage{multirow}
\usepackage{amsmath}	
\usepackage{amssymb}	
\usepackage{gensymb}
\usepackage{colortbl}

\usepackage{array,etoolbox}
\preto\tabular{\setcounter{magicrownumbers}{0}}
\newcounter{magicrownumbers}
\def\rownumber{}

\def\apjl{ApJ}                
\def\apjs{ApJS}               
\def\ao{Appl.~Opt.}           
\def\aap{A\&A}                

\def\src {NGC 4559 X7}
\def\ulx2 {NGC 4559 ULX-2}
\def\swift{{\it Swift/XRT}}
\def\nus{{\it NuSTAR}}
\def\chandra{{\it Chandra}}
\def\xmm{{\it XMM-Newton}}

\def\lum{erg s$^{-1}$}
\def\msun{M$_{\odot}$}

\usepackage[normalem]{ulem}

\title[Flaring activity in NGC 4559 X7]{The rare X-ray flaring activity of the Ultraluminous X-ray source NGC 4559 X7}

\author[F. Pintore et al.]{Fabio Pintore$^{1,2}$\thanks{E-mail: fabio.pintore@inaf.it}, S. Motta$^3$, C. Pinto$^{2}$, M.G. Bernardini$^{3}$, G. Rodriguez-Castillo$^{2,4}$,  \newauthor R. Salvaterra$^1$, G. L. Israel$^{4}$, P. Esposito$^{5}$, E. Ambrosi$^2$, C. Salvaggio$^{3, 6}$, L. Zampieri$^7$, \newauthor A. Wolter$^3$  \\
$^1$ INAF -- IASF Milano, Via E. Bassini 15, 20133 Milano, Italy; \\
$^2$ INAF -- IASF Palermo, Via U. La Malfa 153, 90146 Palermo, Italy; \\
$^3$ INAF, Osservatorio Astronomico di Brera, via Brera 28, 20121 Milano, Italy; \\
$^4$ INAF - Osservatorio astronomico di Roma, Via Frascati 44, I-00040, Monteporzio Catone, Italy; \\ 
$^5$ Scuola Universitaria Superiore IUSS Pavia, Piazza della Vittoria 15, 27100 Pavia, Italy; \\
$^6$ Dipartimento di Fisica, Universit\`a degli Studi di Milano-Bicocca, Piazza della Scienza 3, I-20126 Milano, Italy; \\
$^7$ INAF-Osservatorio Astronomico di Padova, Vicolo dell'Osservatorio 5, I-35122 Padova, Italy \\
}

\pubyear{2020}

\begin{document}

\maketitle

\begin{abstract}
Ultraluminous X-ray sources are considered amongst the most extremely accreting objects in the local Universe. The recent discoveries of pulsating neutron stars in ULXs  strengthened the scenario of highly super-Eddington accretion mechanisms on stellar mass compact objects. In this work, we present the first long-term light curve of the source \src\ using all the available \swift, \xmm, \chandra\ and \nus\ data. Thanks to the high quality 2019 \xmm\ and \nus\ observations, we investigated in an unprecedented way the spectral and temporal properties of \src. The source displayed flux variations of up to an order of magnitude and an unusual flaring activity. We modelled the spectra from \src \ with a combination of two thermal components, testing also the addition of a further high energy cut-off powerlaw. We observed a spectral hardening associated with a luminosity increase during the flares, and a spectral softening in the epochs far from the flares.
Narrow absorption and emission lines were also found in the RGS spectra, suggesting the presence of an outflow. Furthermore, we measured hard and (weak) soft lags with magnitudes of a few hundreds of seconds whose origin is possibly be due to the accretion flow. We interpret the source properties in terms of a super-Eddington accretion scenario assuming the compact object is either a light stellar mass black hole or a neutron star.

\end{abstract}
\begin{keywords}
accretion, accretion discs - X-rays: binaries - X-Rays: galaxies - X-rays: individual: NGC 4559 X7.
\end{keywords}

\section{Introduction}

Ultraluminous X-ray sources (ULX) are a class of non nuclear accreting compact objects displaying X-ray luminosities higher than $10^{39}$ erg s$^{-1}$ and up to $10^{41\div42}$ \lum\ \citep[e.g.][]{kaaret17}. The nature of the compact object, in the vast majority of ULXs, is nowadays believed to be of stellar type, i.e. stellar mass black holes (BH; M$_{BH}$ = 5--80 \msun) or neutron stars (NS). In particular, six accreting ULXs hosting pulsating NSs (PULXs), with spin periods in the range 0.5~s -- 20~s have been discovered in recent years \citep{bachetti14,israel16a, carpano18,israel16b,sathyaprakash19,rodriguez19}, suggesting that the NS population in ULXs can be larger than previously expected \citep[see e.g.][]{middleton17,pintore17,koliopanos17,walton18,king20}. It is believed that, because of the transient behaviour of the observed pulsations in PULXs, many NSs in ULXs can remain undetected and can be unveiled only through indirect evidences as, for instance, the presence of cyclotron absorption lines (as in the case of  M51 ULX-8; \citealt{brightman18,middleton19}) or a flux switch-off due to propeller phase \citep[e.g.][]{tsygankov16,earnshaw18}. A further indication of a NS accretor may be the presence of a hard spectrum, a common characteristic of all PULXs, and which may be associated to an accretion column above the NS surface \citep[e.g.][]{pintore17,walton18,walton20}. 

\begin{table}
\label{log}
\scalebox{0.75}{\begin{minipage}{24cm}
\begin{tabular}{@{\makebox[1.3em][l]{\rownumber\space}} | llccr}
\hline
Instr. & Obs.ID & Start & Stop & Exp. \\
 & & \multicolumn{2}{c}{[YYYY-MM-DD hh:mm:ss (TT)]} & [ks] 
\gdef\rownumber{\stepcounter{magicrownumbers}\arabic{magicrownumbers}} \\
\hline
\chandra & 2026 & 2001-01-14T16:24:39 & 2001-01-14T19:27:35 & 9.4 \\
\chandra & 2027 & 2001-06-04T01:14:29 & 2001-06-04T04:43:23 & 10.7 \\
\chandra & 2686 & 2002-03-14T05:13:59 & 2002-03-14T06:34:19 & 3.0 \\
\xmm & 0152170501 & 2003-05-27T03:07:30 & 2003-05-27T14:19:33 & 42.2 \\
\swift & 00032249001 & 2012-01-14T15:53:31 & 2012-01-14T20:41:16 & 3.9 \\
\swift & 00032249003 & 2012-01-16T01:01:11 & 2012-01-16T06:17:57 & 4.2 \\
\swift & 00032249004 & 2012-01-18T01:33:23 & 2012-01-18T19:21:55 & 3.1 \\
\swift & 00032249005 & 2012-01-20T04:35:58 & 2012-01-20T12:58:55 & 3.8 \\
\swift & 00576064000 & 2013-10-28T05:46:26 & 2013-10-28T07:38:28 & 3.5 \\
\swift & 00032249006 & 2014-02-27T18:39:59 & 2014-03-03T11:21:54 & 4.9 \\
\swift & 00032249007 & 2014-03-05T23:46:07 & 2014-03-05T23:59:55 & 0.8 \\
\xmm & 0842340201 & 2019-06-16T19:03:44 & 2019-06-17T15:06:27 & 74.3 \\
\swift & 00088825001 & 2019-06-17T22:40:58 & 2019-06-17T23:08:53 & 1.7 \\
\nus & 30501004002 & 2019-06-17T21:31:09 & 2019-06-20T04:01:09 & 94.9 \\
\swift & 00032249008 & 2019-12-12T00:46:13 & 2019-12-12T01:09:52 & 1.4 \\
\swift & 00032249009 & 2019-12-19T12:50:14 & 2019-12-19T13:12:53 & 1.4 \\
\swift & 00032249011 & 2020-01-02T08:32:43 & 2020-01-02T08:47:53 & 0.9 \\
\swift & 00032249012 & 2020-01-09T06:04:44 & 2020-01-09T06:21:52 & 1.0 \\
\swift & 00032249013 & 2020-01-16T00:40:59 & 2020-01-16T00:50:53 & 0.6 \\
\swift & 00032249014 & 2020-01-23T15:43:19 & 2020-01-23T16:05:53 & 1.3 \\
\swift & 00032249015 & 2020-01-30T16:42:51 & 2020-01-30T17:04:52 & 1.3 \\
\swift & 00032249016 & 2020-02-06T12:51:41 & 2020-02-06T13:14:53 & 1.4 \\
\swift & 00032249017 & 2020-02-13T13:49:35 & 2020-02-13T14:08:54 & 1.2 \\
\swift & 00032249018 & 2020-02-20T00:24:36 & 2020-02-20T02:08:52 & 1.4 \\
\swift & 00032249019 & 2020-02-27T12:21:16 & 2020-02-27T12:43:52 & 1.4 \\
\swift & 00032249020 & 2020-03-05T11:39:38 & 2020-03-05T12:01:52 & 1.3 \\
\swift & 00032249021 & 2020-03-12T00:06:31 & 2020-03-12T00:26:52 & 1.2 \\
\swift & 00032249022 & 2020-03-19T10:17:34 & 2020-03-19T10:40:54 & 1.4 \\
\swift & 00032249023 & 2020-03-26T12:51:57 & 2020-03-26T13:14:54 & 1.4 \\
\swift & 00032249024 & 2020-04-02T12:11:47 & 2020-04-02T12:33:54 & 1.3 \\
\swift & 00032249025 & 2020-04-09T11:36:37 & 2020-04-09T11:53:52 & 1.0 \\
\swift & 00032249026 & 2020-04-16T07:50:00 & 2020-04-16T08:10:54 & 1.2 \\
\swift & 00032249027 & 2020-04-23T02:09:17 & 2020-04-23T02:32:54 & 1.4 \\
\swift & 00032249028 & 2020-04-30T11:11:28 & 2020-04-30T11:33:52 & 1.3 \\
\swift & 00032249029 & 2020-05-07T12:05:07 & 2020-05-07T12:28:54 & 1.4 \\
\swift & 00032249030 & 2020-05-14T11:29:05 & 2020-05-14T11:45:53 & 1.0 \\
\swift & 00032249031 & 2020-05-21T12:24:35 & 2020-05-21T12:42:53 & 1.1 \\
\swift & 00035479001 & 2006-01-02T18:56:44 & 2006-01-03T04:47:56 & 5.3 \\
\swift & 00032249032 & 2020-05-28T11:29:03 & 2020-05-28T11:51:52 & 1.4 \\
\swift & 00032249033 & 2020-06-04T07:44:35 & 2020-06-04T08:11:53 & 1.6 \\
\swift & 00032249034 & 2020-06-11T07:06:17 & 2020-06-11T07:28:53 & 1.4 \\
\swift & 00032249035 & 2020-06-18T04:46:38 & 2020-06-18T05:09:52 & 1.4 \\
\swift & 00032249036 & 2020-06-25T07:14:56 & 2020-06-25T07:38:53 & 1.4 \\
\swift & 00032249037 & 2020-07-02T05:00:02 & 2020-07-02T05:20:53 & 1.2 \\
\swift & 00032249038 & 2020-07-09T01:00:01 & 2020-07-09T01:23:53 & 1.4 \\
\swift & 00032249039 & 2020-07-16T03:41:41 & 2020-07-16T04:00:52 & 1.1 \\
\swift & 00032249040 & 2020-07-23T20:20:22 & 2020-07-23T20:46:51 & 1.6 \\
\swift & 00032249041 & 2020-07-30T07:01:53 & 2020-07-30T07:25:55 & 1.4 \\
\swift & 00032249042 & 2020-08-06T12:36:18 & 2020-08-06T13:01:52 & 1.5 \\
\swift & 00032249043 & 2020-10-28T13:48:32 & 2020-10-28T14:12:52 & 1.5 \\
\swift & 00032249044 & 2020-11-04T16:15:16 & 2020-11-04T16:37:53 & 1.4 \\
\swift & 00032249045 & 2020-11-11T05:57:36 & 2020-11-11T17:30:53 & 1.5 \\
\swift & 00032249046 & 2020-11-18T03:39:42 & 2020-11-18T04:01:53 & 1.3 \\
\swift & 00032249047 & 2020-11-25T06:09:00 & 2020-11-25T06:33:52 & 1.5 \\
\swift & 00032249048 & 2021-01-24T17:41:56 & 2021-01-24T18:09:53 & 1.7 \\
\swift & 00032249049 & 2021-01-31T09:03:13 & 2021-01-31T22:09:54 & 1.6 \\
\swift & 00032249050 & 2021-02-07T16:35:44 & 2021-02-07T16:44:54 & 0.5 \\
\swift & 00032249051 & 2021-02-14T15:27:42 & 2021-02-14T20:36:53 & 1.5 \\
\swift & 00032249052 & 2021-02-21T14:53:45 & 2021-02-21T22:52:53 & 1.6 \\
\swift & 00032249053 & 2021-02-28T04:44:06 & 2021-02-28T20:26:52 & 1.3 \\
\swift & 00032249054 & 2021-03-07T10:06:21 & 2021-03-07T13:53:52 & 0.5 \\
\swift & 00032249055 & 2021-03-14T11:03:30 & 2021-03-14T19:15:51 & 1.5 \\
\hline
\end{tabular}
\end{minipage}}
\caption{Log of the observations used in this work.}
\end{table}

The assumption that the accretor is a stellar-mass compact object implies super-Eddington accretion, where a certain degree of geometrical beaming may be taken into account to explain the observed high X-ray luminosity in ULXs. Magneto-hydrodynamical simulations showed that, in case of super-Eddington accretion, strong optically thick outflows in the form of winds can be radiatively ejected by the accretion disc \citep[e.g.][]{poutanen07,ohsuga11,takeuchi14}, from radii smaller than the so called spherization radius (R$_{sph}\propto \dot{m_0}$, where $\dot{m_0}$ is the accretion rate in Eddington rate unit; e.g. \citealt{poutanen07}). Observational evidence in favour of powerful outflows/winds were identified in some ULXs in which blue-shifted ($z\sim0.2\text{c}$) absorption lines have been detected \citep[e.g.][]{pinto16,pinto17,kosec18b}. In addition, such winds are believed to be the main cause of the time lags between soft and hard X-ray photons observed in some ULXs, such as NGC 55 ULX-1 \citep{pinto17}, NGC 5408 X-1 \citep[e.g.][]{heil10,demarco13} and NGC 1313 X-1 \citep{kara20}. Such lags, conventionally labelled as negative lags (or soft lags, because the soft photons lag the hard ones), are observed at low frequencies ($\sim$0.1--10 mHz) with a magnitude of tens to thousands of seconds, and are thought to originate from the propagation of hard photons through a very extended optically thick medium, likely associated with the base of the outflows. Because of the generally low ULX count-rate, time lags appear to be preferentially observed in sources with high short-term variability. The presence of flaring activity was also observed in epochs where lags were found, as in the case of the ULX NGC 1313 X-1 \citep{kara20}. However, flares as well as heart-beat variability are not so common in ULXs and only an handful of sources showed them (e.g. NGC 7456 ULX-1, \citealt{pintore20}; NGC 253 ULX-1, \citealt{barnard10}; NGC 6946 ULX-3, \citealt{earnshaw19b}; NGC 247 ULX-1, Pinto et al. in prep.; 4XMM J111816.0-324910 in NGC 3621, \citealt{motta20}).

In this work, we report on the source NGC 4559 X7 (also know as RX J123551+27561; X7 hereafter), a ULX in the galaxy NGC 4559, which showed a peculiar flaring activity during X-ray observations taken with \xmm\ and \nus\ in 2019. 
NGC 4559 is a spiral galaxy historically assumed to be at a distance of $\sim$10 Mpc (\citealt{tully88,sanders03}; although the lower distance limit is $\sim$7 Mpc, \citealt{sorce14}). The galaxy hosts two ULXs \citep[X7 and X10; e.g.][]{soria04} that have been poorly studied in the past. Only a few short \chandra\ and \xmm\ observations of NGC 4559 are available in the archives, during which the two ULXs showed extremely high 0.3--10 keV luminosity ($>10^{40}$ \lum), if a distance to the source of 10 Mpc is assumed. X7 lies in the outskirts of the galaxy (RA = 12h 35m 51.71s, Dec = +27d 56m 04.1s; \citealt{swartz11}), inside a region rich of OB-type stars with low metallicity ($0.2<\text{Z}/\text{Z}_{\odot}<0.4$),  where \citet{soria04} identified its possible optical counterpart as a blue supergiant of 20 \msun\ and an age of $\sim$10 Myr.

\begin{figure*}
\center
\includegraphics[width=17.5cm]{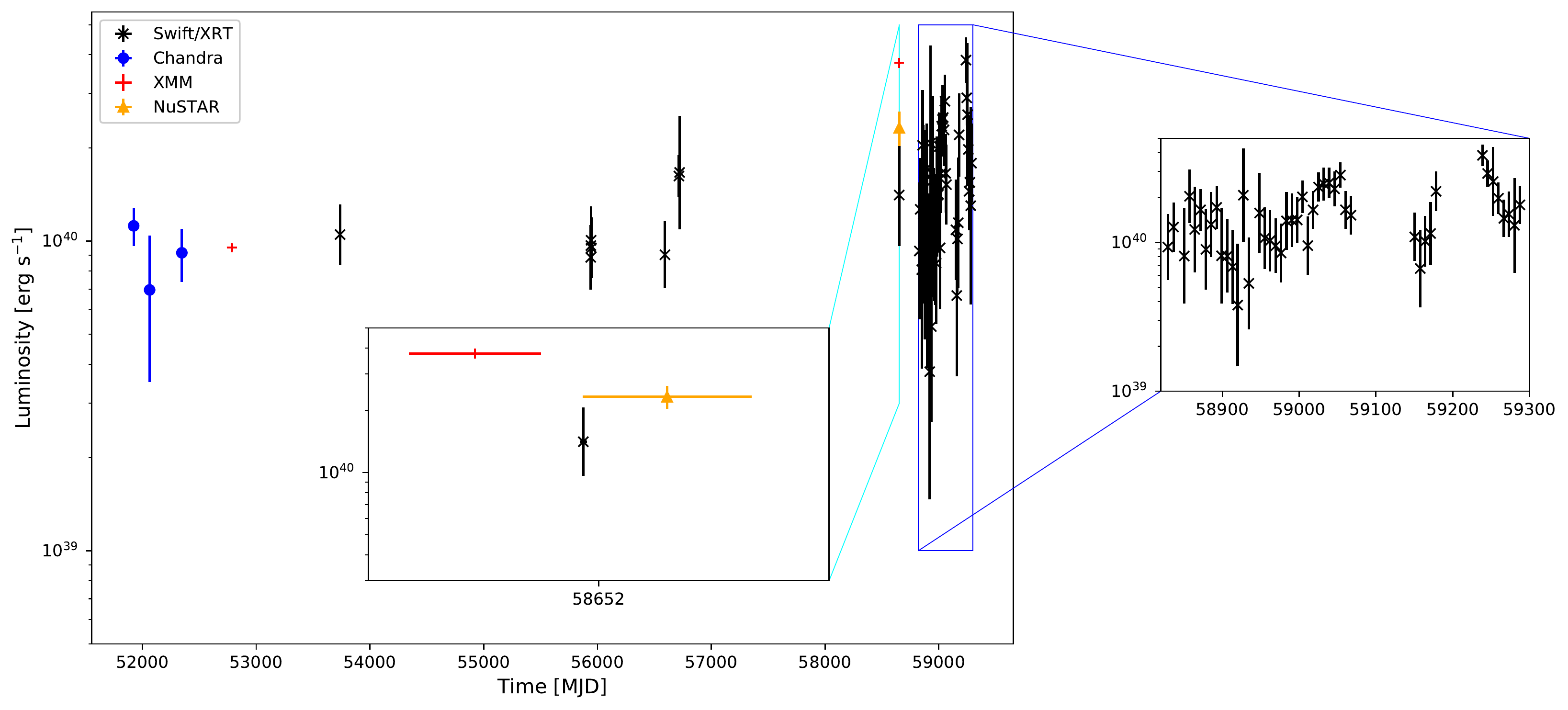}
   \caption{Long-term lightcurve of X7 in the 0.3--10 keV energy band obtained from the \xmm, \nus, \chandra\ and \swift\ observations. The luminosities  (not corrected for absorption) refer to the average values for all the observations and are calculated assuming a distance of 10 Mpc. The source luminosities in the \swift\ observations are estimated by adopting an absorbed powerlaw model (see text for details).}
   \label{lcall}
\end{figure*}

\section{data reduction}
\label{data_reduction}

\vspace{0.1cm}
\noindent {\bf XMM-Newton.} \xmm\ observed NGC 4559 on 27 May 2003 for $\sim$42 ks. Then our group obtained a new $\sim$75 ks-long observation, which was taken on 16 June 2019 (PI: F. Pintore; Table~\ref{log}). We reduced the two observations with {\sc SAS} v18.0.0. We used data from both EPIC-pn and EPIC-MOS (1 and 2) cameras: events were selected considering {\sc pattern}$\leq$4 (i.e. single and double-pixel events) for the PN, and {\sc pattern}$\leq$12 (i.e. single- and multiple-pixel events) for the MOS. We extracted source and background events from circular regions with radii of 30'' and 60'', respectively. The photon times of arrival (ToAs) were converted to the barycenter of the Solar System with the task \textsc{BARYCEN}, using the best \chandra\ coordinates (RA=12h 35m 51.71s, Dec=+27d 56m 04.1s; \citealt{swartz11}) of the target.

\noindent Both observations were only marginally affected by high background epochs, which were excluded from the analysis. This resulted in a net exposure time of $\sim$$35$ks and $\sim$$65$ks for the observations taken in 2003 and 2019, respectively.

\noindent Source and background spectra were grouped, with the FTOOLS {\it grppha}, in order to accumulate at least 25 counts per energy bin, and we applied the SAS task {\sc epiclccorr} to all light curves.

\noindent We also used data from the Reflection Grating Spectrometer (RGS; \citealt{denHerder2001}) of the 2019 observation, where X7 was at center of the field of view (FoV). The RGS data were reduced with the \textsc{rgsproc} task, which produces calibrated event files, spectra, response matrices, and 1D images.
Following the standard procedures, we filtered the RGS data for solar flares using the background light curve from the RGS CCD number 9 (corresponding to $\lesssim7.5$\,{\AA} or $\gtrsim1.7$\,keV). The background light curves were binned at a 100~s time-resolution and all the time bins with a count rate above 0.2 cts s$^{-1}$ were rejected. We used the same GTI for both the RGS 1 and 2, and obtained an exposure time of 71.2\,ks for each detector. The net exposure time is slightly longer than the EPIC one owing to the lower RGS background contamination. 

\vspace{0.1cm}
\noindent {\bf NUSTAR.} The \nus\ observation (Obs.ID: 30501004002) was taken about 6.4 h after the 2019 \xmm\ observation (Table~\ref{log}). The total exposure time of this observation is $\sim$95 ks.
We processed and reduced the data with the standard {\sc nupipeline} task of the \nus\  {\it  Data Analysis Software} v1.3.0 ({\sc NUSTARDAS}) in the HEASOFT package version 6.25, adopting standard filtering and corrections. Spectra were extracted from circular regions with radius 60$''$ for both source and background. Spectra were then rebinned to accumulate at least 25 counts per energy bin.

\vspace{0.1cm}
\noindent {\bf Chandra.} We analyzed the \chandra\ observations of 14 January and 4 June 2001, and 14 March 2002, with exposures of 9.7 ks, 11 ks and 3 ks, respectively (Table~\ref{log}). We reduced the data with {\sc ciao} v.4.9 and calibration CALDB v.4.7.6.  
We selected source and background events from circular regions with radii 3'' and 15'', respectively.
The source spectra were obtained with the task {\sc specextract} of the {\sc ciao} package, which produces appropriate response and auxiliary files. 
\noindent Source and background spectra were grouped with {\it grppha} in order to accumulate at least 25 counts per energy bin. 

\vspace{0.1cm}
\noindent {\bf Swift.} The {\it Neil Gehrels Swift Observatory} (hereafter \textit{Swift}) observed the galaxy NGC 4559 for 56 times between January 2006 and March 2021 (Table~\ref{log}). Most observations were taken in PC mode with an average exposure time of about 1--2 ks each. We reduced the data with the {\sc xrtpipeline} tool and we extracted source and background events from circular regions of radii 40'' and 60'' (centered at the source position and in a source-free region of the CCD), respectively. We used the \swift\ observations only to construct a long-term light curve, as the data quality is not good enough to allow us to perform high quality spectral analysis of individual or stacked pointing. We estimated the source flux in each observation by fitting an absorbed powerlaw model with photon index and column density fixed at 2 and $1.5\times10^{21}$ cm$^{-2}$, respectively, and letting only the powerlaw normalization to vary. The choice of this reference model was adopted because of the poor quality of the spectra.

\begin{figure*}
\center
\includegraphics[width=9.1cm]{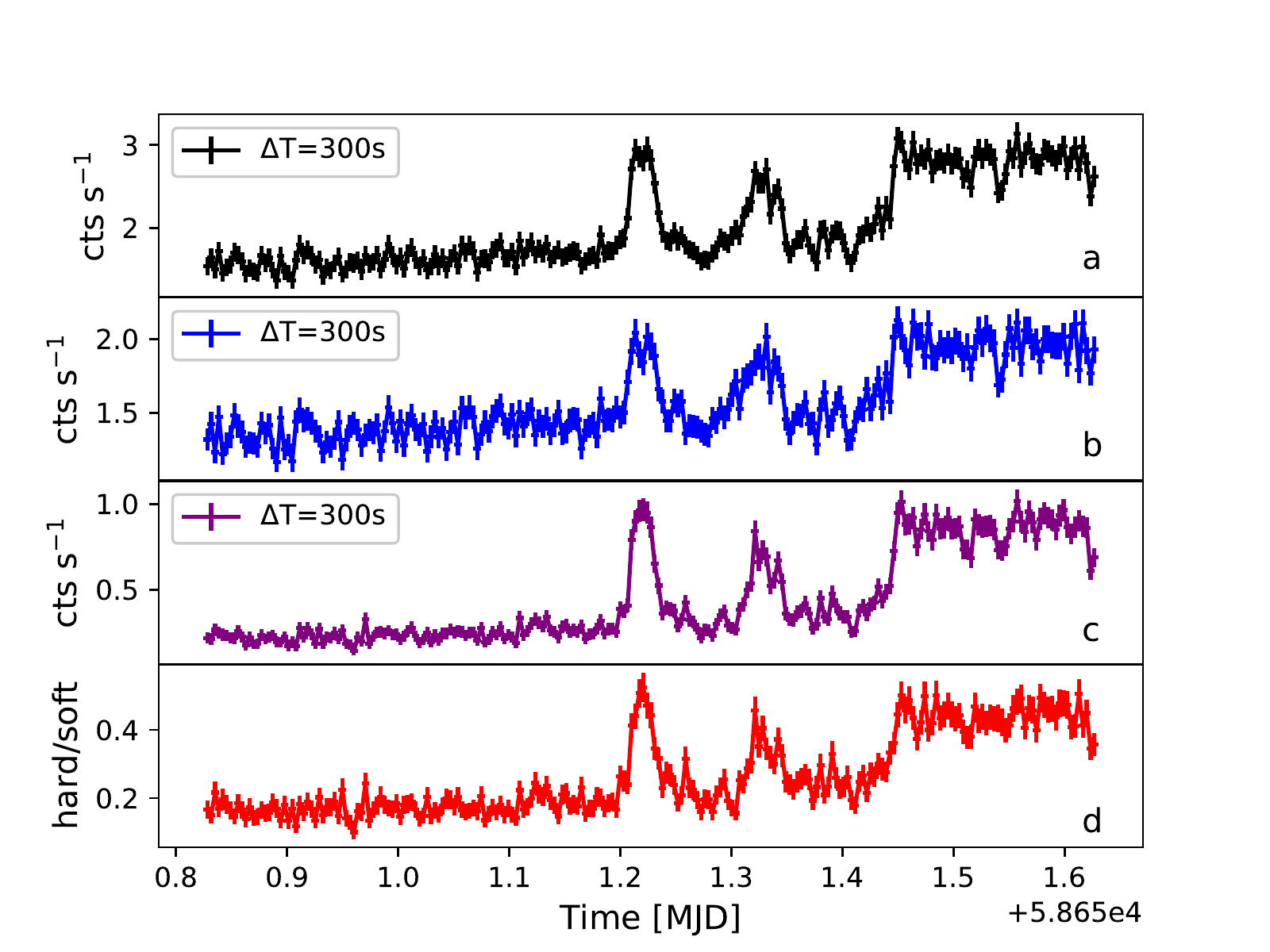}
\hspace{-0.75cm}
\includegraphics[width=9.1cm]{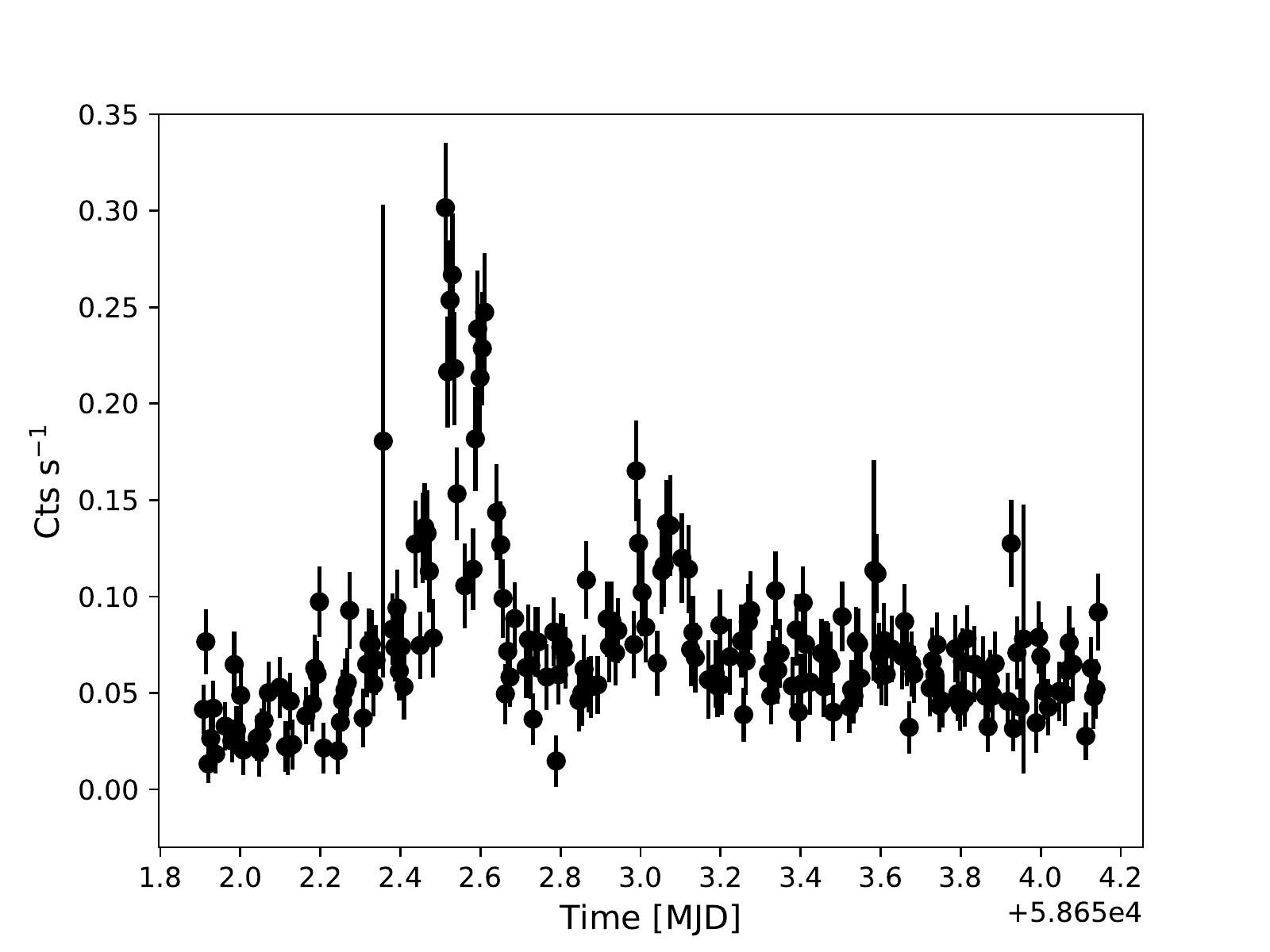}
   \caption{Left: EPIC-pn+MOS1/2 light-curves of X7 in the 0.3-10 keV (panel a), and 0.3-2.0 keV (panel b) and 2.0-10 keV (panel c), binned at 300s. In panel d, we also show the hardness ratio (hard band vs soft band) along the observation. There is a clear hardening of the source during the high flux epochs and the flares. Right: background-subtracted \nus/FMPA+FMPB lightcurve binned at 500s. All the curves start on MJD 58650.}  
        \label{lculx1}
\end{figure*}

\begin{figure}
\center
\includegraphics[width=8.9cm]{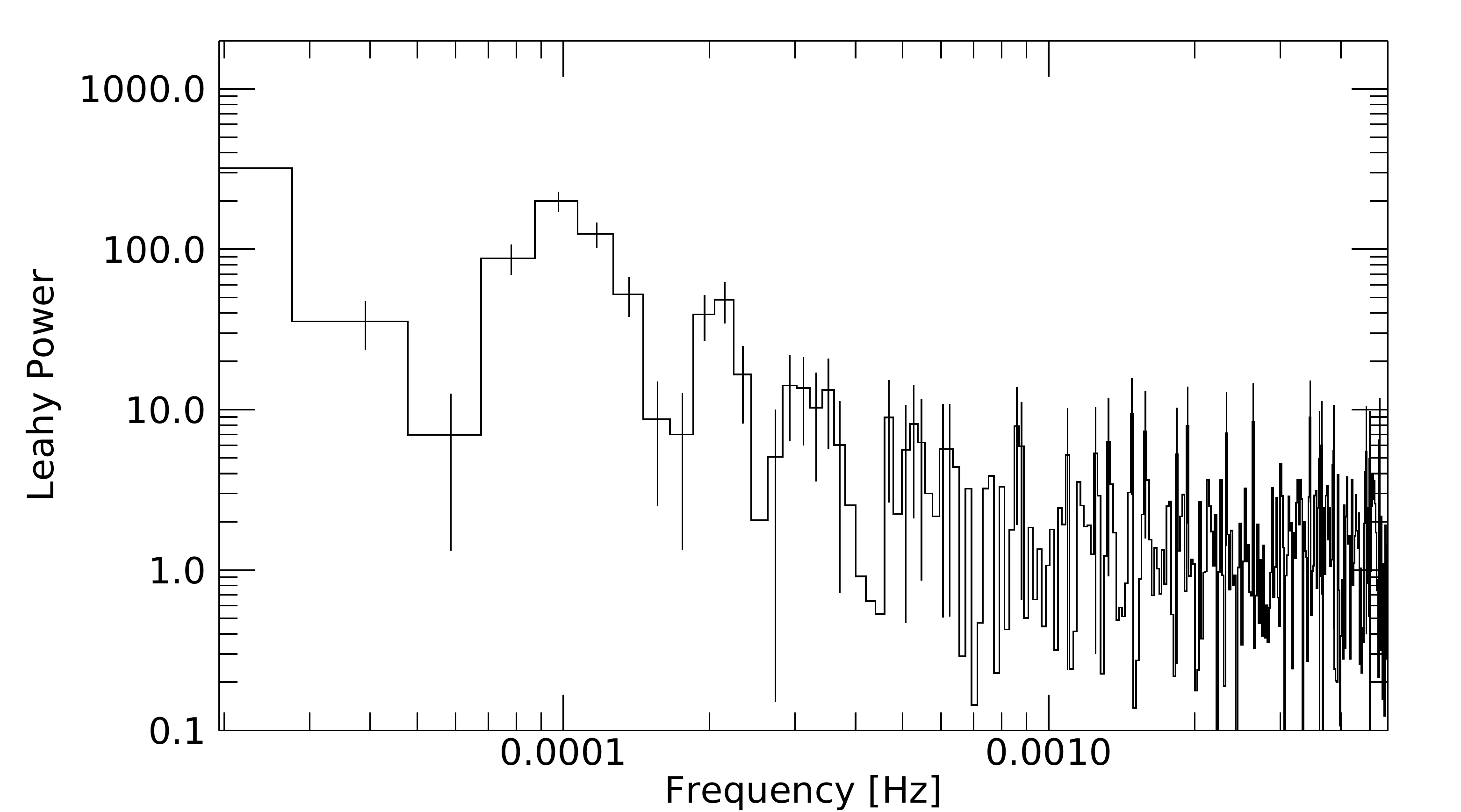}
   \caption{Power density spectrum calculated combining the data from all the \xmm\ EPIC cameras. }\label{fig:xmm_pds}
        \label{lcpower}
\end{figure}

\begin{figure*}
\center
\includegraphics[width=8.5cm]{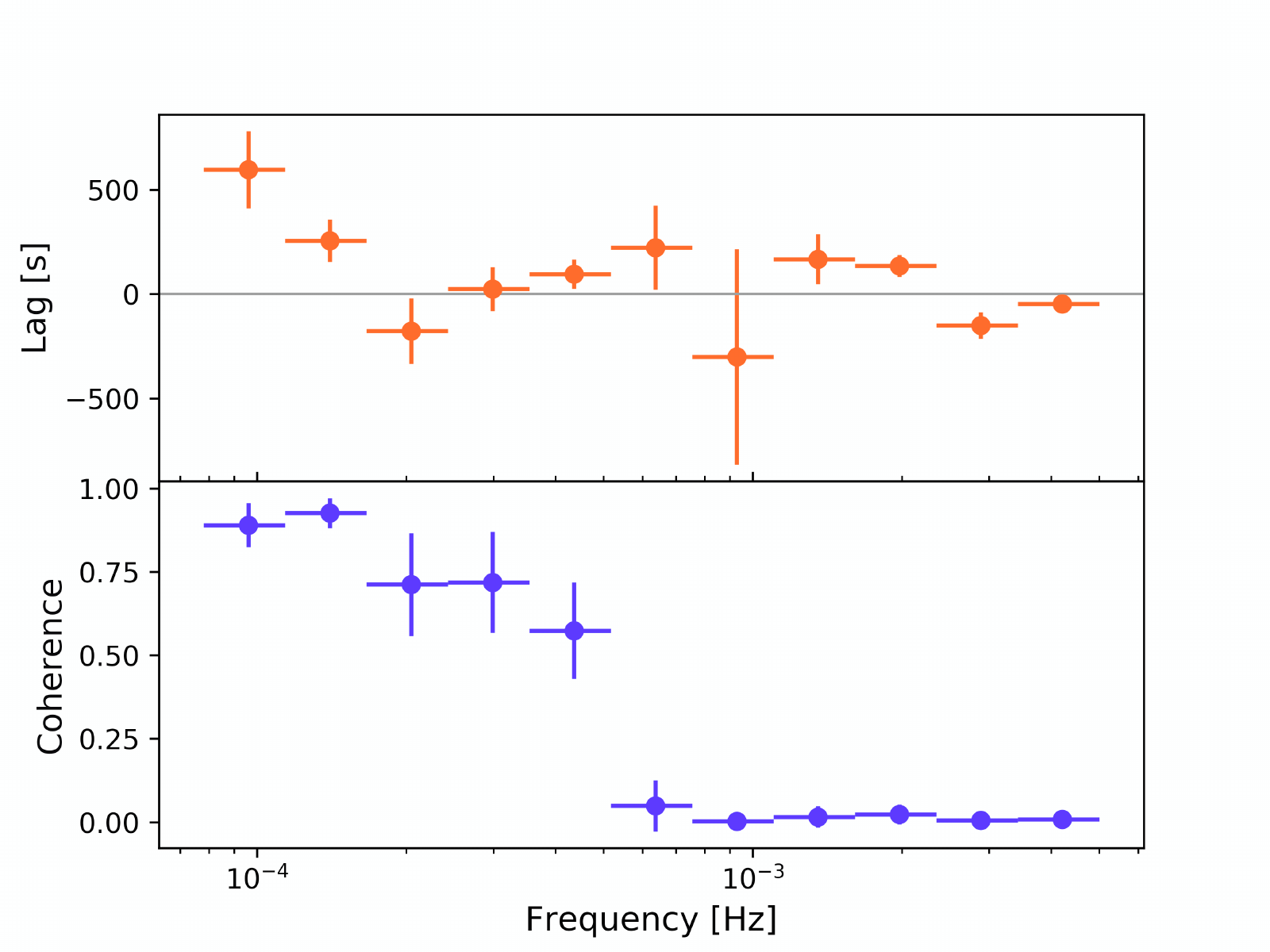}
\includegraphics[width=8.5cm]{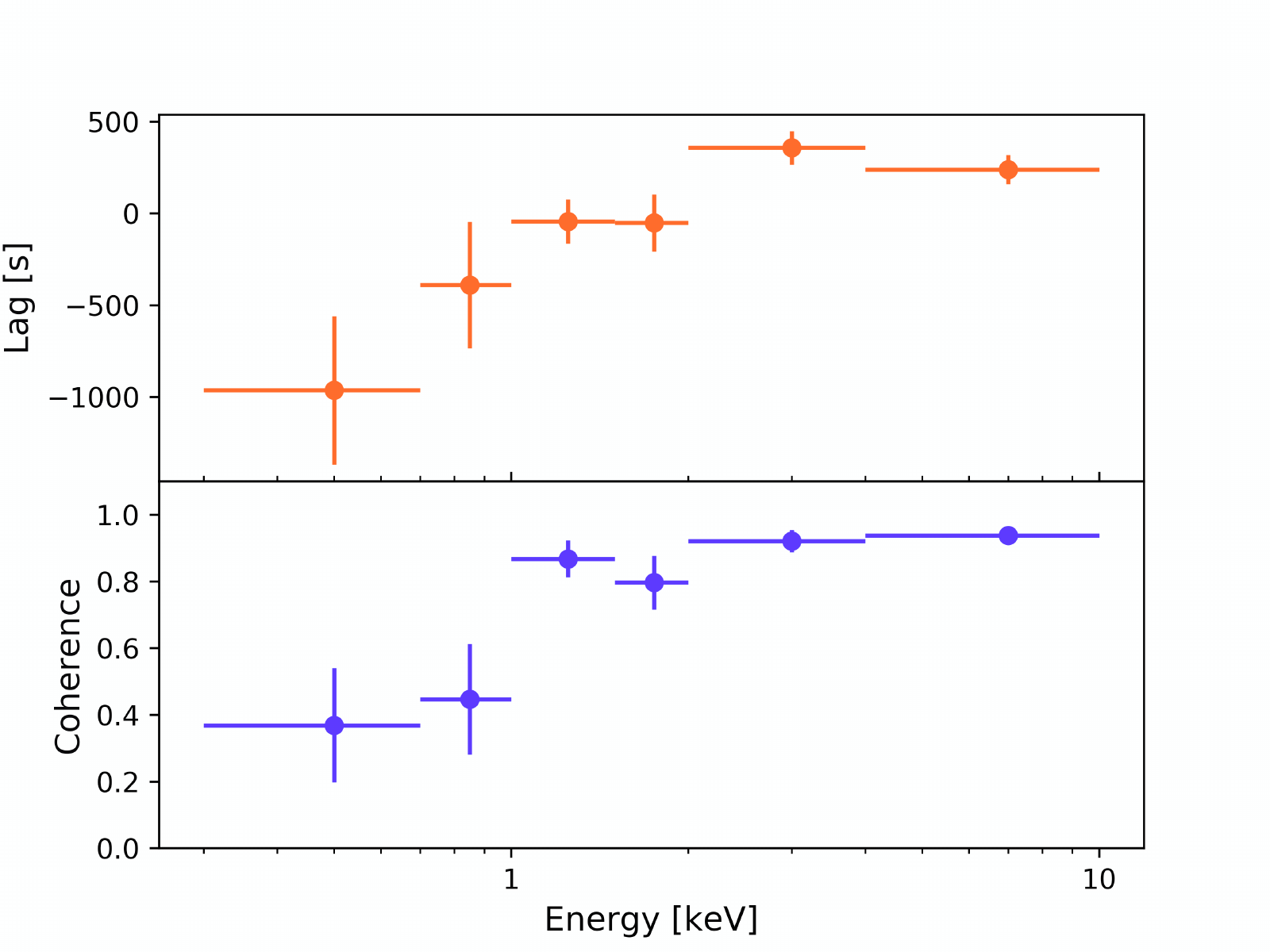}
\includegraphics[width=8.5cm]{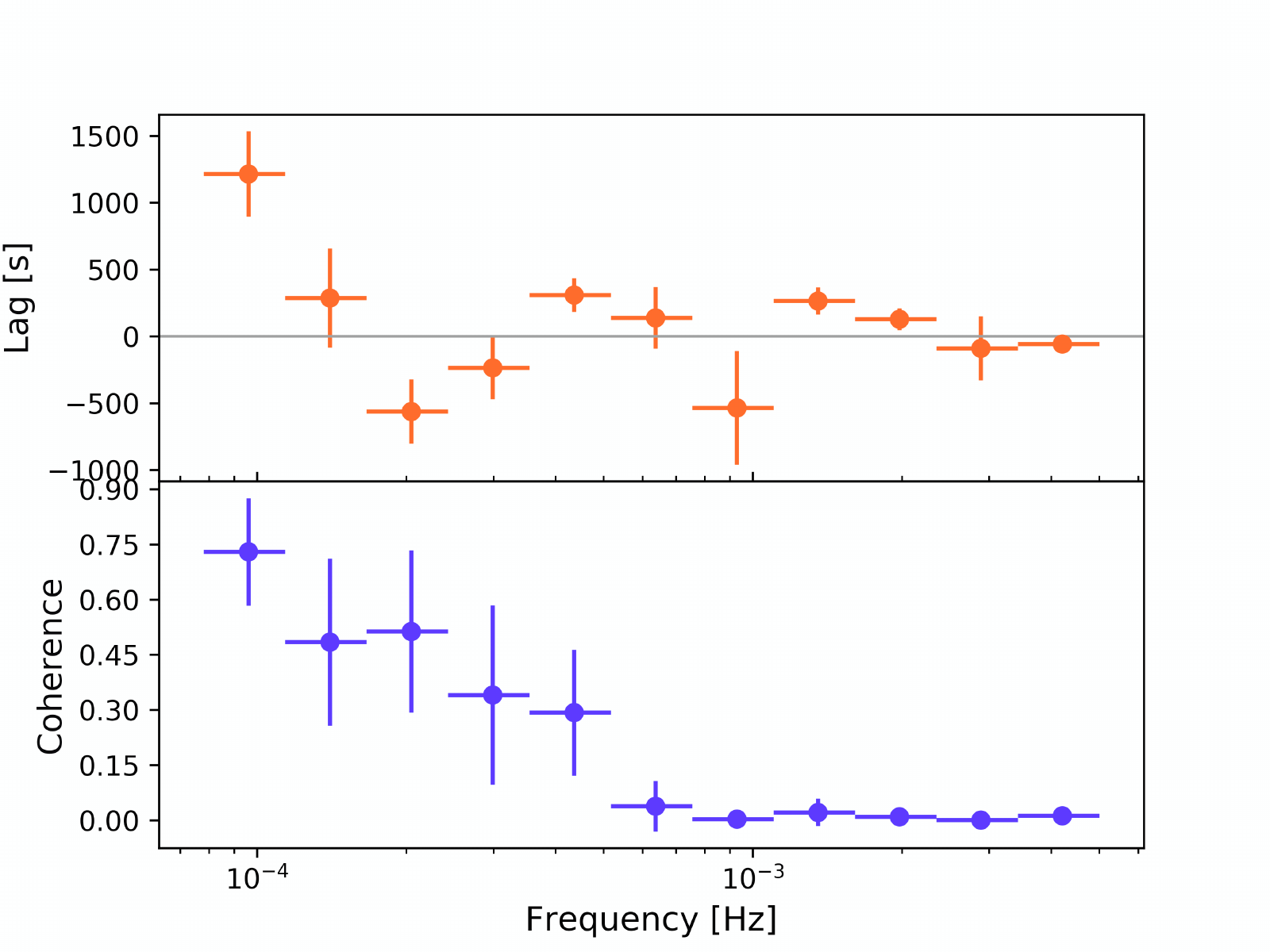}
\vspace{-0.2cm}
 \caption{Time lag of X7 as a function of the frequency between the bands 0.3-2.0 keV and 2.0-10 keV (top-left) and the lag spectrum in the $(0.7-1.6)\times10^{-4}$ Hz frequency range (top-right). Bottom: lag vs frequency for the energy ranges 0.3-1.0 keV and 2.0-10 keV. \newline According to the standard convention, positive lags imply the hard band lagging the soft band.}
\label{lagulx1}
\end{figure*}

\section{Long and short term variability}

The long-term X-ray light curve shows that X7 varies in flux by up to a factor of $\sim$5-6. The observed source flux peaked during the 2019 \xmm\ observation of our X-ray campaign. \swift\ and \nus\ pointed at X7 a few hours later, finding the source at a mean flux a factor of $\sim3$ lower than during the \xmm\ observation. This indicates that flux variations in X7 may happen on a few-hours time-scales (Fig.~\ref{lcall}). 

Remarkably, the 2019 \xmm\ and \nus\ observations caught a period of flaring activity of X7, never seen in any of the previous X-ray observations of the source. In particular, we found two flares in the \xmm\ observation, at MJD $\sim$ 58651.22 and 58651.33, and at least four (or possibly five) flares in the \nus\ observation, at MJD $\sim$ 58652.46, 58652.52, 58652.61 and 58652.99 (Fig.~\ref{lculx1}). Furthermore, in the last $\sim15$ ks of the \xmm\ observation (i.e. from MJD 58651.44), we observed a high flux phase following the flares, which plateaued at a flux level compatible with that at the peak of the flares (see Fig.~\ref{lculx1}). During the high-flux phase, the source flux increased by more a than a factor of $\sim$4 with respect to the persistent pre-flare emission, implying that the luminosity of X7 can vary by over almost an order of magnitude.

The flares show a variable duration of $\sim$5--10 ks and, based on the \xmm\ observation, their shape is mildly energy-dependent (Fig.~\ref{lculx1}; we stacked the EPIC-pn and MOS1-2 cameras to obtain the shown light curve). The first flare in the \xmm\ curve presents a double peak at energies $<2$ keV, but appears to be single-peaked at higher energies. Despite the lower signal-to-noise ratio of the \nus\ data, we note that in these data the flares seem to be grouped: at least three flares occurring in fast succession can be identified around MJD 58652.6, and two flares around MJD 58653. The limited count statistics of the data prevents us to perform a more in-depth analysis, although a visual inspection suggests that the flares do not show any obvious repeating pattern. 

We also note the existence of ``dip''-like events during the high flux epochs in the last 15~ks of the \xmm\ observation. Such features last for $\sim1$ks and resemble the dips seen in NGC 55 ULX-1 \citep{stobbart04}, as they appear stronger at higher energies. Unfortunately, we cannot determine if they are really ULX-like dips or rather a storm of not resolved flares.

Finally, the \swift\ observations taken in 2020--2021 seems to show a possible (super-)orbital variability of X7 (see zoom in Fig.~\ref{lcall}), which is even more evident in the 1.5--10 keV energy band. We performed a Lomb-Scargle analysis on these data only (i.e. we excluded all the scanty observations taken prior to 2020) and we found a possible best period of $\sim190$ d. Unfortunately, the number of covered periods is still poor, which makes not possible yet to confirm robustly such a variability. Further continuous \swift\ monitoring are needed.

\section{2019 XMM-Newton and NuSTAR observations}

\subsection{Timing}

We created three light-curves of X7 by combining events collected with all the EPIC cameras in the energy band 0.3--10\,keV. Then we binned the data with 140~s (Nyquist frequency $\rm{Ny} = 0.0071$\,Hz).
In order to reach the best possible frequency resolution ($\delta \nu = 1/T \approx 3\times 10^{-5}$\,Hz), we calculated one single power density spectrum (PDS) from the entire light-curve. The resulting PDS is shown in Fig.\,\ref{fig:xmm_pds}. The \xmm \ power density spectrum appears featureless, characterised by low-amplitude red noise. A low-significance narrow feature appears at frequency $\approx 10^{-4}$ Hz, which can be ascribed to the flares visible in the light curve. 
We investigate the presence of variability on shorter timescales as well (from a few tens to a few hundreds seconds). The resulting power density spectrum appears completely dominated by the instrumental noise, and no other feature is significantly detected above it.

We also produced an average PDS from the \nus\ observation. A typical \nus\ observation features a large number of data segments separated by orbital gaps, which prevent from exploring via Fourier analysis the presence of long-term variability on timescales longer than the typical data segment ($\sim 1000$s). Furthermore, the count rate of X7 as seen by \nus\ is low (between 0.05 and 0.25 counts s$^{-1}$). Therefore, for both the FMPA and FMPB instruments on-board \nus, we only explored the short timescales by producing PDS over data stretches of variable length (between 130s and 1050s), which we averaged into one final PDS to increase the signal-to-noise ratio. Similarly to the \xmm\ case, the resulting PDS are dominated by instrumental noise, and no significant feature is detected. 

\subsection{Search for pulsations}

We searched for coherent signals in the 2019 \xmm/EPIC-pn light curves of X7 in the 0.3--10\,keV energy range (about 72000 counts) by adopting the recipe outlined in \citet{israel96}. We firstly considered the effects of signal smearing introduced by the possible presence of a strong first period derivative component $\dot{P}$, as seen in many PULXs (as an example see \citealt{israel16a}). We corrected the photon arrival times by a factor $-\frac{1}{2}\frac{\dot{P}}{P}\,t^2$ for a grid of about 2000 points in the range 4$\times$10$^{-6}<|\frac{\dot{P}}{P}\,($s$^{-1})|<$5$\times$10$^{-12}$  (see \citealt{rodriguez19} for details). No significant peak was found and 3$\sigma$ upper limits to the pulsed fraction (PF), defined as the semi-amplitude of the sinusoid divided by the average count rate, were derived for the source. 
In the best cases, we obtained upper limits around 5\%--8\% in the 150\,ms--150\,s range. We then searched for pulsations applying orbital corrections for orbits with periods in the range $\sim$\,2 h -- 10 d, and composed by a NS and a companion star mass ranging from $\sim$\,0.01 up to 100\,M$_\odot$. No convincing pulsations were detected, but only a marginal detection of a signal at $\sim$\,1.7\,s, with PF $\sim$\,8\% and significance $\sim$\,3$\sigma$, for an orbital period of about 2.8\,h and a companion star of mass $\sim$0.2\,M$_\odot$. However, we note that such a companion mass value would be in contrast with the blue supergiant counterpart identified by \citet{soria04}.

To investigate the 1.7\,s candidate signal, we used its orbital parameters and all the available observations with enough statistics to reveal a $\sim$\,8\% PF signal. We searched in the 2003 \xmm\ observation and in our \nus\ (see Table \ref{log}) observation, which are the only two observations that could at least marginally detect such a low PF signal. We applied the corrections corresponding to the orbital parameters, but the 1.7\,s signal candidate could not be confirmed. We do report on this candidate signal in future observations, with better counting statistics, which could allow us to convert its marginal significance to a robust detection.

\subsection{Time lags}

The appearance of flaring activity in X7 prompted us to search for time-lags between different energy bands, similar to those  observed in the ULXs NGC 5408 X-1, NGC 55 ULX1 and NGC 1313 X-1 \citep{demarco13,pinto17,kara20}. We performed the analysis focusing only on the EPIC-pn data of the 2019 \xmm\ observation, to avoid stacking data from different instruments that may introduce fake variability. In order to do so, we created background-subtracted light curves in the 0.3--2 keV and 2--10 keV energy bands, where the source counting statistics are comparable. 

We first performed a cross-correlation between the two light curves in the time domain by using a non-mean subtracted discrete cross-correlation function (CCF, \citealt{Band97}).
We used light-curves binned to a time resolution of  $\Delta T=20$s (in order to exclude all the white noise at high frequencies) and choose the 2--10 keV range as reference energy band. We calculated the CCF value for a series of time delays $k\Delta T$ and we defined the lag as the time delay that corresponds to the global maximum of the circular CCF versus time delay, located by fitting an asymmetric Gaussian model to the CCF versus time delay. The uncertainties on the CCF have been derived by applying a flux-randomization method \citep{Peterson98}.
We found marginal indications of a positive lag (hard lag) of $94\pm44$s (1$\sigma$ uncertainty), which, according to our convention, implies that the hard photons lag the soft photons.

We further investigated how the hard lag depends on the frequency. Following the recipe described in \citet{uttley14}, we calculated the Fast Fourier transform of the light curves in the 0.3--2 keV and 2--10 keV energy bands, binned with $\Delta T=100$s, and we calculated the CCF. The latter was then averaged over 5 intervals, each of them containing segments of 128 time bins. Lags were then grouped in equally spatially logarithmic bins of frequency.

High coherence\footnote{It is estimated, adopting the prescription of \citet{uttley14}, as $\gamma^2(\nu_j) = (|C_{XY}(\nu_j)|^2 - n^2) / (P_X(\nu_j)~P_Y(\nu_j))$, where $C_{XY}(\nu_j)$ is the cross-spectrum at the $j$-frequency, $n^2$ is the bias term and $P_{(X/Y)}$ are the PDS of the X and Y bands.} between the two bands are observed for frequencies lower than $\sim$$10^{-3}$ Hz, while above this threshold the coherence dramatically drops to values consistent with zero (Fig.~\ref{lagulx1}/top-left). A hard lag in the frequency range $(0.7-1.6)\times10^{-4}$ Hz is seen, in a frequency region where the coherence is also high. The global significance of the lag (estimated by summing the significance of each lag point) in the $(0.7-1.6)\times10^{-4}$ Hz frequency range is higher than $5\sigma$. This result still confirms that, on long time-scales, the hard band lags the soft one for a weighted averaged value of $335\text{s}\pm89$s, compatible within uncertainties with the value derived from the CCF. The frequency range at which the lag is found is compatible with the time interval between the two observed flares, which implies that the lag is likely mainly driven by such temporal features. 

We verified the robustness of the positive time lag by performing a Monte Carlo simulation of 1000 light-curves having the same variability observed in the EPIC-pn 0.3--10 keV energy band, rescaled to the mean count-rate found in the 0.3--2.0 keV and 2.0--10 keV bands. For each couple of simulated 0.3--2.0 keV and 2.0--10 keV light curves, we estimated the frequency-dependent lags and we inferred that a hard lag in the range $(0.7-1.6)\times10^{-4}$ Hz can be observed with a significance higher than $5\sigma$ only 3 times over a total of 1000, i.e. a probability of 0.997 (hence $\geq3\sigma$), that the observed lags in the 2019 \xmm\ observation are not due to chance coincidence.

\begin{figure*}
\center
\includegraphics[width=12cm]{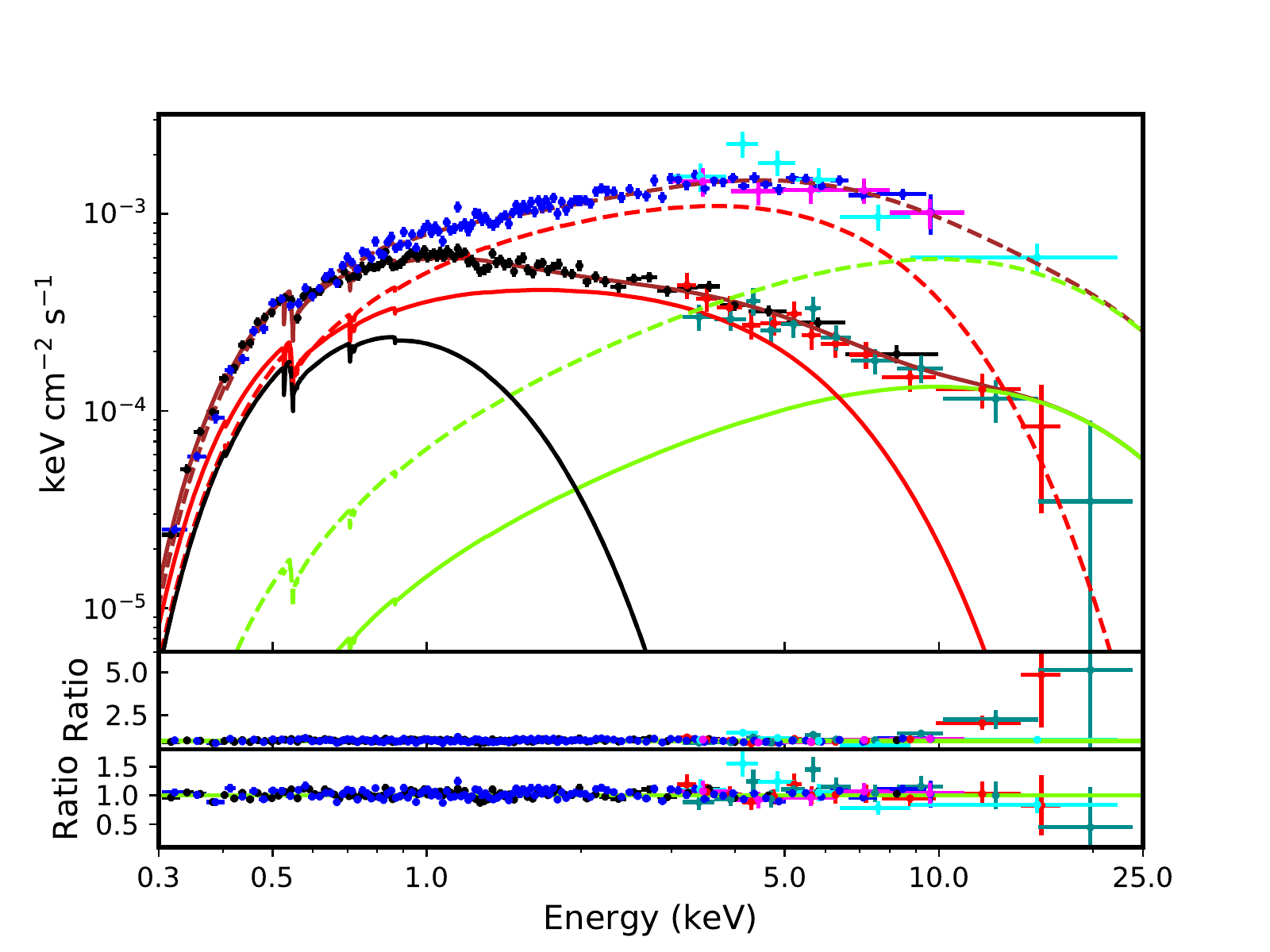}
   \caption{Unfolded {\it persistent} EPIC-pn (black points) + \nus\ (red, dark cyan points) spectrum, and {\it flare} EPIC-pn (blue points)+ \nus\ (purple and cyan points) spectrum. Black, red and green solid lines indicate the {\sc diskbb}, {\sc diskPbb} and {\sc cutoffpl} best-fit models, respectively, for the {\it persistent} spectrum, while the brown solid line is the total model; we use the same color coding for the dashed lines, which refer to best-fit models of the {\it flare} spectrum. Middle panel: residuals of the fit with an absorbed {\sc diskbb+diskPbb} model. Bottom panel: residuals of the best-fit with an absorbed {\sc diskbb+diskPbb+cutoffpl} model. Spectra have been rebinned for display purposes only.}  
        \label{spec_flare_no_flare}
\end{figure*}

We then extracted the lag-energy spectrum of the observed data in the frequency range $(0.7-1.6)\times10^{-4}$ Hz. For each energy bin, we calculated the lag using the 0.3--10 keV band as reference, from which we removed each band of interest. This is shown in Fig.~\ref{lagulx1}/top-right. The coherence clearly increases towards higher energies. At energies below 2 keV the lags are quite unconstrained although they suggest a soft lag, while above this energy the lags are significantly positive.

Besides the hard lag, we also note the hint ($1\sigma$ significance) of a soft lag at frequencies around $2\times10^{-4}$ Hz. However, when using the energy bands 0.3--1.0 keV and 2--10 keV, we found that the detection of the soft lag increases ($\sim2.5\sigma$ significance). The lag energy spectrum at these frequencies is shown in Fig.~\ref{lagulx1}-bottom. 

\subsection{Spectral analysis}
\label{spec_average}
The hardness ratio between the energy bands 0.3--2.0 keV and 2.0--10 keV of the 2019 \xmm\ observation clearly indicates a marked spectral variability along the observation, which in particular shows evidence of a hardening during the flaring activity (Fig.~\ref{lculx1}-left, bottom panel). We investigated the spectral variability in both the \xmm\ and \nus\ 2019 observations, extracting a {\it persistent} and {\it flare} spectrum from both the \xmm\ and the \nus\ observations.

In {\it XMM}, the {\it persistent} spectrum was extracted over the first part of the observation, where no flares were detectable ($\sim28$ ks). The  {\it flare} spectrum was extracted by considering only the times when the count-rate was higher than 1.1 cts s$^{-1}$ in the 0.3--10 keV EPIC-pn lightcurve, which resulted in a net exposure of 18 ks. In \nus, we instead selected the {\it persistent} and {\it flare} epochs manually. The {\it flare} spectrum was extracted by choosing the latter as time intervals where the count-rate was significantly higher than the average one. 
Even though we adopted a different approach for \xmm\ and \nus, we highlight that the {\it flare} and {\it persistent} spectra obtained with the two satellites are well over-imposed above 3 keV. 

\begin{table}
\scalebox{0.89}{\begin{minipage}{24cm}
\begin{tabular}{lllll}
\hline
Model & Component &\multicolumn{2}{c}{2019 obs.} & Previous obs. \\
 &  &\multicolumn{1}{c}{No flare} &\multicolumn{1}{c}{Flare} &  \\
\hline
 {\sc TBabs} & nH ($10^{22}$) & \multicolumn{2}{c}{$0.11^{+0.01}_{-0.01}$} & - \\
{\sc nthcomp} & kT$_{seed}$ (keV) & \multicolumn{2}{c}{$0.13^{+0.01}_{-0.01}$} & - \\
 & $\Gamma$ & $2.47^{+0.04}_{-0.03}$ & $1.75^{+0.02}_{-0.02}$ & - \\
 & E$_{cut}$ &$3.7^{+1.6}_{-0.8}$ & $2.06^{+0.1}_{-0.1}$ & - \\
 & norm (10$^{-4}$) & $7.0^{+0.2}_{-0.2}$ & $9.4^{+0.3}_{-0.2}$ & - \\
\\
 & $\chi^2/dof$ & \multicolumn{2}{c}{1752.03/1804} & \\
 \hline
 \hline
{\sc TBabs} & nH ($10^{22}$) & \multicolumn{2}{c}{$0.138^{+0.009}_{-0.008}$} & $0.16^{+0.03}_{-0.03}$\\
{\sc diskbb} & Tin (keV) & \multicolumn{2}{c}{$0.28^{+0.02}_{-0.02}$} & $0.20^{+0.03}_{-0.03}$\\
 & norm & \multicolumn{2}{c}{$8.3^{+3.0}_{-2.1}$} & $20^{+24}_{-10}$\\
{\sc diskPbb} & Tin (keV) & $2.0^{+0.2}_{-0.1}$ & $2.4^{+0.1}_{-0.1}$ & $2.1^{+0.5}_{-0.4}$\\
 & p & $0.500^{+0.004}_{0}$ & $0.616^{+0.009}_{-0.009}$ & $0.54^{+0.06}_{-0.03}$\\
 & norm (10$^{-3}$) & $1.1^{+0.4}_{-0.3}$ & $4.5^{+1.0}_{-0.9}$ & $0.6^{+1.3}_{-0.4}$\\
 \\
 & $\chi^2/dof$ & \multicolumn{2}{c}{1777.1/1803} & 492.3/465\\
 \hline
 \hline
 {\sc TBabs} & nH ($10^{22}$) & \multicolumn{2}{c}{$0.15^{+0.01}_{-0.01}$} & - \\
{\sc diskbb} & Tin (keV) & \multicolumn{2}{c}{$0.26^{+0.01}_{-0.02}$} & - \\
 & norm & \multicolumn{2}{c}{$13^{+6}_{-4}$} & - \\
{\sc diskPbb} & Tin (keV) & $1.3^{+0.2}_{-0.1}$ & $1.9^{+0.4}_{-0.3}$ & - \\
 & p & $0.5^{+0.016}_{0}$ & $0.6^{+0.03}_{-0.01}$ & - \\
 & norm (10$^{-3}$) & $6.5^{+3.6}_{-2.5}$ & $7.5^{+8.0}_{-3.0}$ & - \\
{\sc cutoffpl} & $\Gamma$ & \multicolumn{2}{c}{0.59 (frozen)} & - \\
 & E$_{cut}$ & \multicolumn{2}{c}{7 (frozen)} & - \\
 & norm (10$^{-5}$) & $2.2^{+0.5}_{-0.5}$ & $10^{+5}_{-7}$ & - \\
\\
 & $\chi^2/dof$ & \multicolumn{2}{c}{1750.3/1801} & \\
 \hline
 \hline
\end{tabular}
\end{minipage}}
\caption{Best-fit spectral parameters. Errors are at 90$\%$ uncertainty for each parameter of interest.}
\label{spec_all}
\end{table}

We simultaneously analyzed the \xmm+\nus\ spectra of the two epochs in the range 0.3--30 keV, using {\sc Xspec} v.12.10.1. We adopted {\sc tbabs} to model the absorption, assuming no variations of the absorption column density along the line of sight during the two epochs (i.e. we linked the parameter between the spectra). As typically found for ULXs, the X7 spectra cannot be described by simple models such as an absorbed single {\sc powerlaw} or a {\sc diskbb} ($\chi^2_{\nu}>3$). The data are well-modelled by either a single Comptonization model ({\sc nthcomp} in {\sc Xspec}, \citealt{zdiarski96}; $\chi^2_{\nu}=0.97/1804$) or a combination of thermal models. A model in the form {\sc tbabs $\times$(diskbb+diskPbb\footnote{The radial dependence of the disc temperature of the {\sc diskPbb} goes as $r^{-p}$, with $p=0.75$ for a standard disc and $p=0.5$ for a slim disc.})} provides a statistically acceptable fit, with ($\chi^2_{\nu}=0.99/1803$). Comparing the {\it flare} and {\it persistent} spectra, the low-energy {\sc diskbb} component is consistent with being constant within 90$\%$ uncertainties, therefore we linked its temperature and normalization to a common value across the spectra.

Even though the resulting best fit is statistically acceptable ($\chi^2_{\nu}=0.99/1803$), we note the presence of a hard excess in the form of structured residuals above 10 keV (see Fig.~\ref{spec_flare_no_flare}, center panel), as in many other PULXs and ULXs observed with \nus\  \citep[e.g.][]{bachetti13,walton20}. To take into account the high energy excess, we added a cut-off power law component ({\sc cutoffpl} in {\sc Xspec}), which has been found previously to be representative of the high energy emission in PULXs. Such a component is usually interpreted as due to the presence of an accretion column above the NS and it appears to be present also in ULXs where the nature of the compact object could not be determined yet \citep{walton18b}. The simple inclusion of a cut-off power-law to the best fit model causes a number of spectral parameters to become unconstrained. Hence, we fixed the spectral parameters of the cut-off power-law to the average values found for the PULXs spectra, i.e. $\Gamma=0.59$ and E$_{cut}=7.1$ keV, and left only the model normalization amongst the spectra free to vary. The final model ({\sc tbabs(diskbb+diskPbb+cutoffpl)}) provided a very good fit ($\chi^2_{\nu}=0.97/1801$; Fig.~\ref{spec_flare_no_flare}-top and bottom panel), which suggests  that during the flaring activity both the {\sc diskPbb} and the {\sc cutoffpl} contributed to the high energy flux increase. All the results are reported in Table~\ref{spec_all}. The measured absorbed 0.3--30 keV fluxes are $(2.08\pm0.03)\times10^{-12}$ erg cm$^2$ s$^{-1}$ and $(5.16\pm0.07)\times10^{-12}$ erg cm$^2$ s$^{-1}$ for the {\it persistent} and {\it flare} spectra, respectively, corresponding to luminosities of $(2.48\pm0.04)\times10^{40}$ erg s$^{-1}$ and $(6.15\pm0.09)\times10^{40}$ erg s$^{-1}$ (for a distance of 10 Mpc).

\subsection{Covariance spectra}

\begin{figure}
\center
\hspace{-0.8cm}
\includegraphics[width=6.2cm, angle=270]{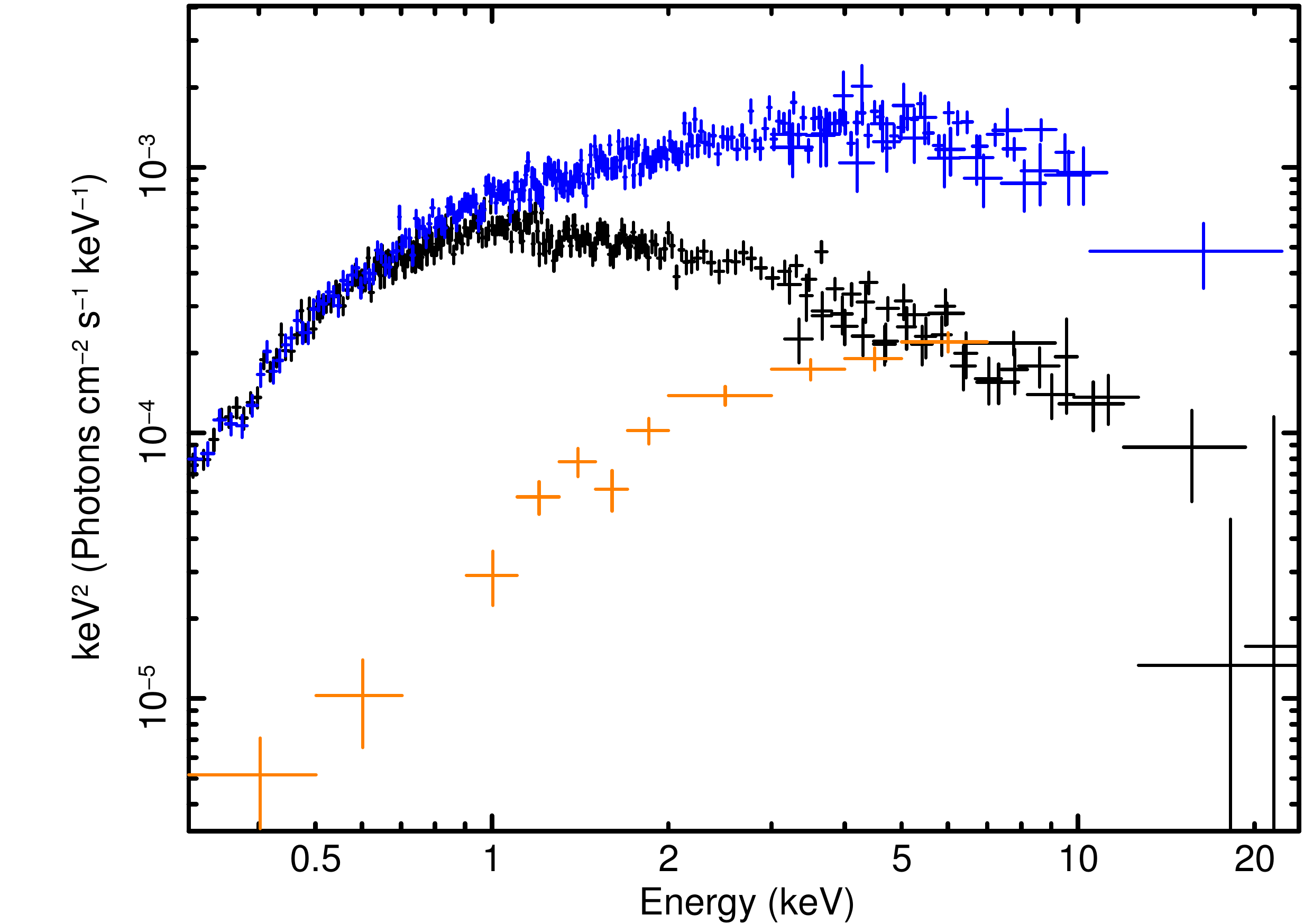}
   \caption{Covariance (orange points), {\it flare} (blue points) and {\it persistent} (black points) spectra, unfolded through a powerlaw of index 0. The covariance spectrum shows a spectral shape similar to the {\it flare} spectrum. Data have been rebinned for display purposes only.}  
        \label{spec_CV_flare_no_flare}
\end{figure}

We extracted the covariance spectrum \citep[see][]{uttley14} from the EPIC-pn data in the frequency range $(0.7-1.6)\times10^{-4}$ Hz. The covariance spectrum measures the spectral behaviour of the short-term variability correlated with a given reference band. It allows to investigate the spectral components that are responsible of the variability. We compared the EPIC-pn covariance spectrum with those of the {\it persistent} and {\it flare} emission, in order to allow us to investigate the correlated spectral variability (Fig.~\ref{spec_CV_flare_no_flare}). At a first glance, the spectral shape of the covariance spectrum is quite similar with the {\it flare} one. 

We fitted the covariance spectrum with the best-fit model {\sc tbabs(diskbb+diskPbb+cutoffpl)} of the {\it flare} spectrum. First, we multiplied it by a constant, which was left as the only free parameter, to account for an overall flux variation. We found that the model does not reproduce the covariance spectral shape correctly. Then, we removed the constant, and left the normalizations of the three spectral components (i.e. {\sc diskbb}, {\sc diskPbb} and {\sc cutoffpl}) free to vary independently. This approach provided a statistically acceptable result ($\chi^2/dof=14.62/9$): we found that the {\sc cutoffpl} model contributes the most to describe the spectrum (normalization of $(4.6\pm0.4)\times10^{-5}$). Instead, the {\sc diskPbb} normalization was almost two orders of magnitude lower  (($1.7\pm1.1$)$\times10^{-4})$ than the values found in either the {\it flare} or {\it persistent} spectra. Furthermore, the {\sc diskbb} normalization was consistent with zero and therefore it could be excluded by the fit. We note that letting free to vary also the {\sc diskPbb} temperature did not improve the quality of the fit. Instead, we found an improvement ($\Delta\chi^2\sim4$ for 1 additional dof) by letting free to vary the photon index of the {\sc cutoffpl} model. With this best-fit, the normalization of the {\sc diskPbb} model became compatible with 0, while the {\sc cutoffpl} photon index and normalization converged to $0.79_{-0.2}^{+0.08}$ and $(6.3_{-1.4}^{+0.6})\times10^{-5}$), respectively. As a final step, we removed also the {\sc diskPbb} component and left free to vary also the cut-off energy of the {\sc cutoffpl} model. The best-fit is highly satisfactory ($\chi^2/dof$ = 8.24/8) and we obtained $\Gamma=0.4\pm0.4$, $E_{\text{cut}}=3.6_{-1.3}^{+3.6}$ keV and $\text{Norm} = (6.5_{-0.8}^{+0.9})\times10^{-5}$. These results indicate that the correlated variability was mainly driven by the {\sc cutoffpl} component.

\subsection{RGS spectra}

\begin{figure}
\centering
\includegraphics[width=6.3cm,angle=-90]{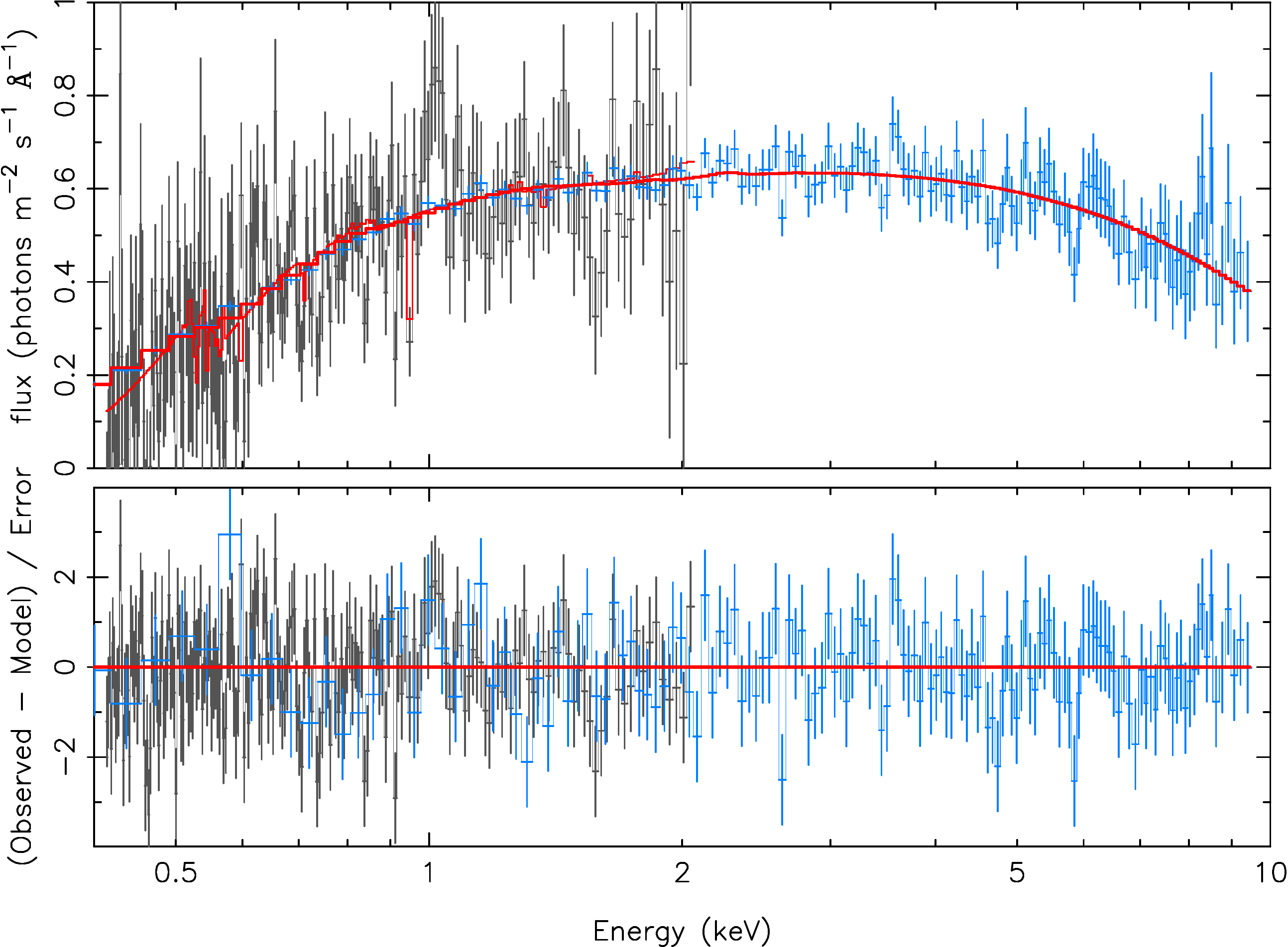}
\vspace{-0.cm}
\caption{\textit{XMM-Newton} EPIC-pn (light blue) and RGS (dark grey) 
         time-averaged spectra of X7 
         from the 2019 \xmm\ observation (top panel). The red solid lines show the best-fit {\scriptsize{SPEX}}
         comptonisation continuum model. The bottom panel shows the corresponding residuals.
         The spectra have been rebinned for displaying purposes.
 \label{fig:spex_plot}}
\end{figure}

The primary goal of the RGS is to resolve and constrain the properties of narrow features that are detected with higher statistics (albeit at lower resolution) in the EPIC spectra in the soft ($<2$ keV) X-ray band. To be consistent with the extraction of the EPIC spectra, we extracted the first-order RGS spectra in a cross-dispersion region of width 1', centred at the emission peak. We extracted the background spectra by selecting photons beyond the $98\%$ of the source point-spread-function and we checked for consistency by comparing the resulting spectrum with the background spectra from blank field data. We stacked the RGS 1 and 2 spectra for plotting purposes with the \textit{rgscombine} task. The time-average stacked RGS spectrum is shown in Fig.\,\ref{fig:spex_plot}  along with the EPIC-pn spectrum.
In order to search for any changes in the narrow spectral features during the source flare events, we extracted flare-resolved spectra from the RGS 1 and 2 by running the \textit{rgsproc} task and using the GTIs that were used to produce the {\it flare} EPIC-pn spectrum. 

The RGS spectra cover a limited energy band, which means that an additional broadband spectrum has to be used in order to constrain the continuum shape, and to avoid systematics in the search for narrow spectral features. Therefore, we simultaneously fitted the RGS and pn spectra because the latter, amongst the \textit{XMM-Newton} detectors, features the highest effective area all the way up to 10 keV. Here we do not use the MOS 1 and 2 spectra for two reasons: 1) their effective area in the Fe-K band is significantly lower than the pn's; 2) their high count rate in the soft band of the RGS (together with that of pn) would wash away the statistical weight from the RGS data, which makes hard to detect the features.

The simultaneous spectral analysis of the RGS and EPIC-pn data is performed using a minimal binning of 1/3 of the line spread function in order to avoid a loss in spectral resolution. We used the {\sc{SPEX}} software {\footnote{http://www.sron.nl/spex}} (\citealt{Kaastra1996}) to carry out the spectral analysis and we adopted the C-statistics \citep{Cash1979}. We analyzed the EPIC-pn spectrum between 0.3--10 keV and the RGS 1 and 2 spectra between 6--30\,{\AA} ($\sim0.4-2$ keV). While in principle the RGS spectra could extend down to 0.3 keV, we note that the background already dominates the X-ray emission below 0.5 keV.

As shown in Section~\ref{spec_average} and in Table~\ref{spec_all}, a simple model consisting of an absorbed Comptonisation component is able to reproduce both the \textit{XMM-Newton} and \textit{NuSTAR} spectra. Therefore, as we have used {\sc{nthcomp}} in the {\sc{Xspec}} modelling, we use the corresponding {\sc{comt}} model in {\sc{SPEX}} to describe the EPIC-pn and RGS spectra in both the time-averaged and flared-resolved approaches. Of course, the {\sc{SPEX~/~comt}} parameters are consistent within the uncertainties with those estimated for {\sc{Xspec~/~nthcomp}}. Unsurprisingly, the column density of the absorbing interstellar medium is also consistent with the values derived within {\sc{Xspec}}. The {\sc{SPEX}} spectral continuum model for the averaged spectra and the corresponding residuals are shown in Fig.~\ref{fig:spex_plot}. 

There are positive residuals in both RGS and pn at 1\,keV and negative residuals on both edges of the energy range, which are similar to those found in many other ULX spectra from \textit{XMM-Newton}, \textit{Chandra}, and \textit{Suzaku}. Such residuals have been interpreted as the evidence for the presence  of winds (see, e.g., \citealt{middleton15b}) and they are often resolved with RGS in rest-frame emission and blue-shifted absorption lines from ionised atomic species (\citealt{pinto16, kosec18b}). We performed a Gaussian line scan over the whole \textit{XMM-Newton} energy band in order to search for the energy centroids of the strongest features. We decided to keep the pn spectrum in the range 0.3--2 keV (avoiding to fit the RGS spectrum alone), in order to minimise systematic effects of the RGS instrumental features although EPIC spectra have a lower spectral resolution. In fact, \citet{kosec18a} have shown that exposures larger than 120\,ks of bright ULXs are typically required to detect significant features in the RGS spectra alone.

\begin{figure}
\centering
\includegraphics[width=8.8cm]{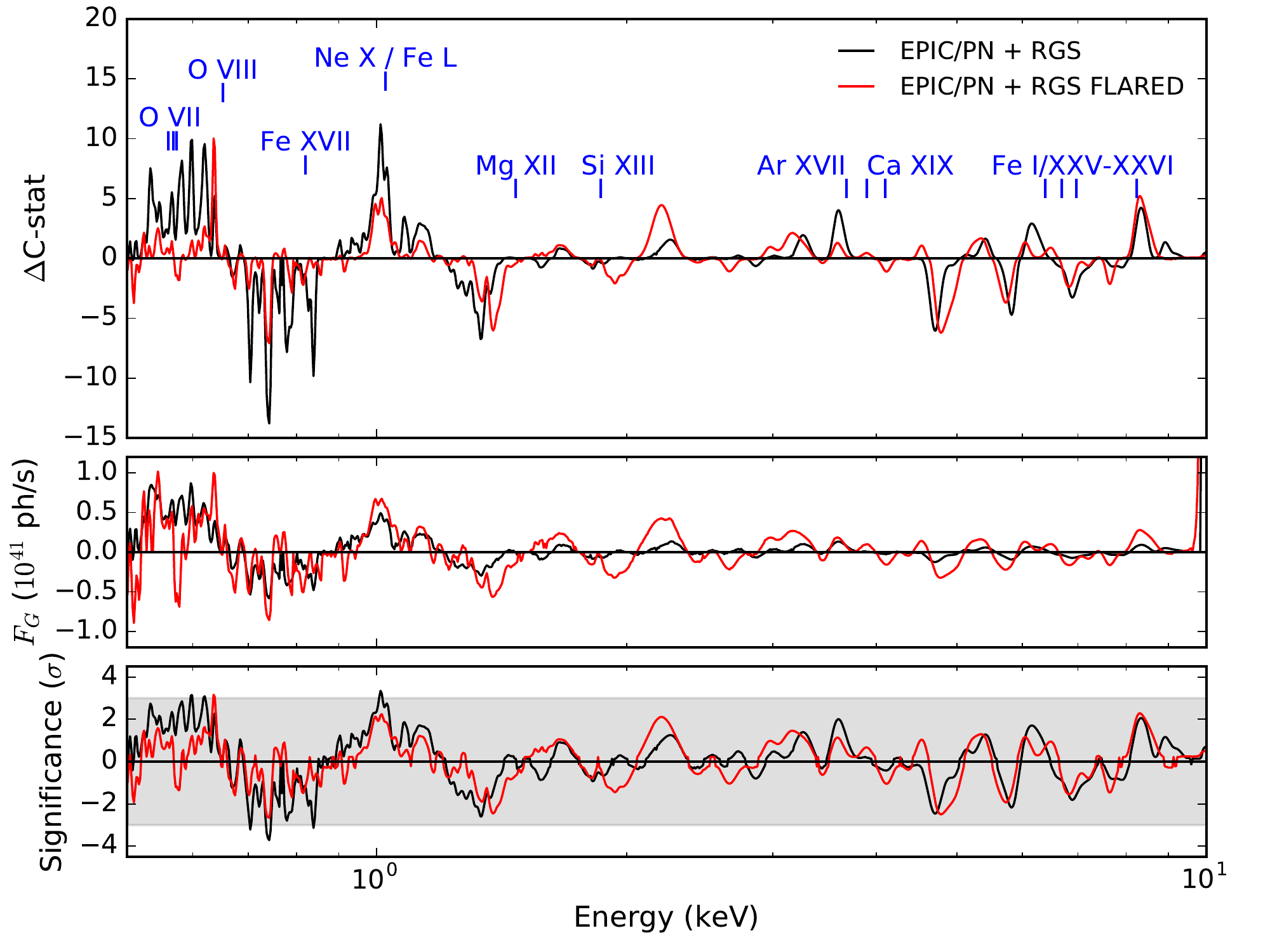}
\vspace{-0.2cm}
\caption{Gaussian line scan performed on the \textit{XMM-Newton} RGS and pn 
   time-averaged (black) and flared-resolved (red) spectra of the 2019 \xmm\ observation. 
   The fit improvement or $\Delta C$-stat (top) and the single trial
   significance (bottom) have been multiplied by the 
   sign of the gaussian normalisation (middle panel) to distinguish between emission 
   (positive flux) and absorption (negative flux) lines.
   The rest-frame energies of the strongest K and L transitions in the X-ray band are labelled.
 \label{fig:spex_scan}}
\end{figure}

Following the approach used in \citet{pinto16}, we searched for narrow spectral features by scanning the RGS and the pn spectra with Gaussian lines, adopting a logarithmic energy grid with energy steps comparable to the RGS and pn resolving power in the 0.5--2 keV and 2--10 keV energy ranges, respectively. We adopted a line width of 500 km s$^{-1}$ 
FWHM (i.e., comparable with the RGS spectral resolution). At each energy bin, we express the single trial significance as the square root of the $\Delta C$-stat. This provides the maximum significance for each line. 


In Fig.~\ref{fig:spex_scan} we show the results of the line scan obtained for the time-averaged (black) and the flare-resolved (red) RGS+pn spectra. The presence of an emission feature at 1 keV (likely due to Ne\,{\scriptsize{X}} or Fe\,L ions) is found, with a possible higher flux during the flaring state. A similar behaviour is seen for the emission-like features around 0.6 keV, which may be due to O\,{\scriptsize{VIII}}. A cluster of absorption features between 0.7--0.8 keV are found and, in previous works, were attributed to blue-shifted O\,{\scriptsize{VII-VIII}} absorption lines. The ions responsible for the 1 keV emission might be the cause of the putative absorption feature seen around 1.6 keV, albeit requiring blue-shift. 

In order to constrain better the line significance, we performed line scan of 1000 EPIC-pn + RGS spectra, simulated with a Monte Carlo approach by adopting the best-fit continuum model of the average spectrum, and we searched for fake lines in the same way as we did for the observed EPIC-pn+RGS spectrum. To address the significance of the three main features, we checked the probability of having two (fake) spectral features yielding $\Delta$C-stat (DC) equal to or above that of the two emission lines found in the real data. This gave a probability of 99\% (i.e. $\sim$2.6$\sigma$) and this was determined by combining the $p$-values of the individual probabilities. A comparable confidence level was obtained by searching simulated spectra for at least 1 spectral feature with DC stronger than that of our most relevant absorption feature at 0.74 keV. The probability of finding individual fake features around $0.5-0.6$ keV and $0.9-1.1$ keV with DC equal or above 10 was even smaller. We therefore believe the real significance of the observed lines to be between $2.6\sigma-3.0\sigma$ each. Ideally one should perform an ad-hoc, time consuming, physical models automatic grid search \citep[e.g.][]{pinto20, kosec18b} but this is not the focus of this work and it will be left for a forthcoming paper.

\section{Archival Chandra and XMM-Newton observations}

\begin{figure}
\center
\hspace{-0.5cm}
\includegraphics[width=6.2cm, angle=270]{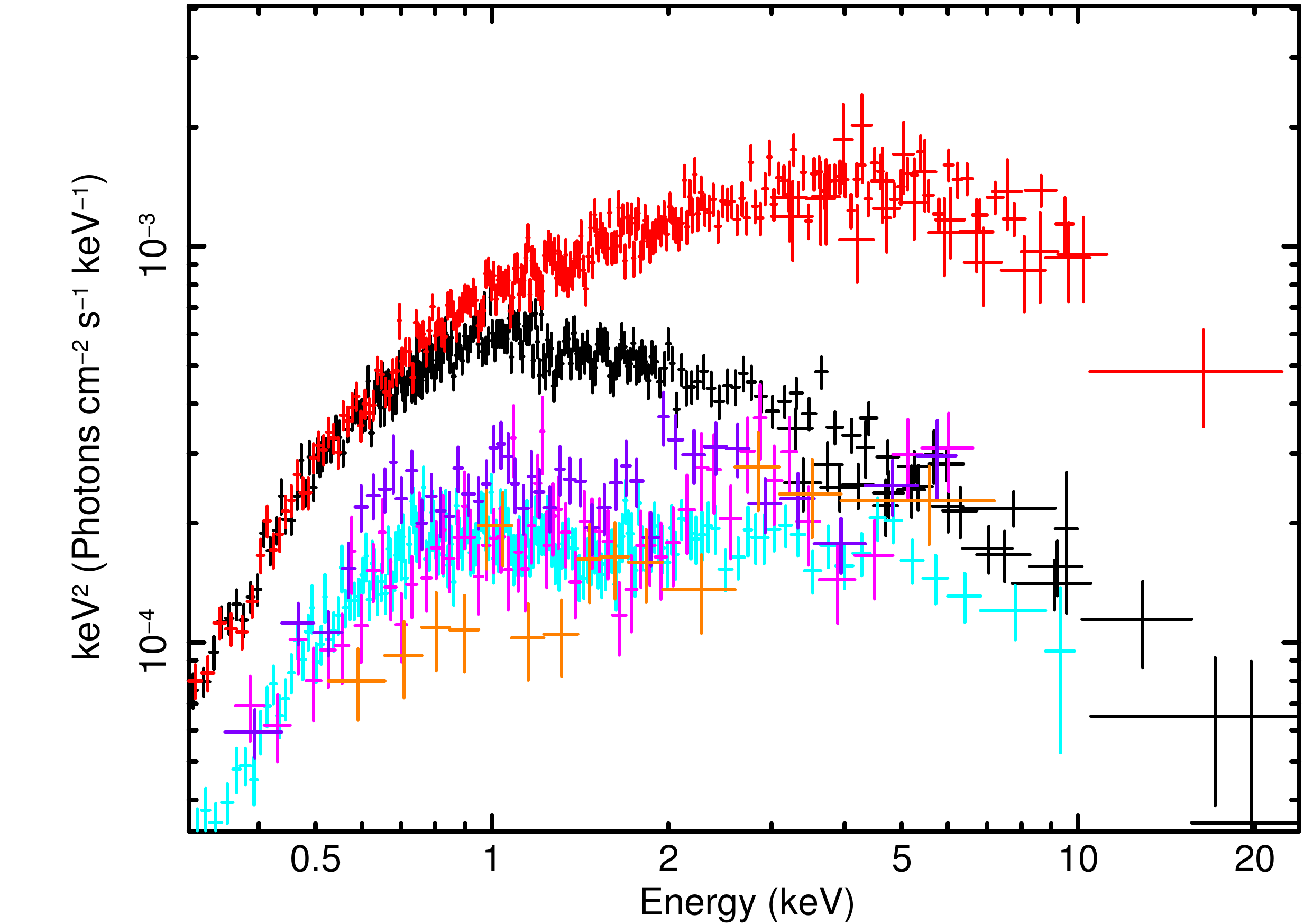}
   \caption{Spectra of all the available high-quality X-ray data unfolded through a powerlaw of photon-index 0. Black and red point refer to the {\it flare} and {\it persistent} spectra of the 2019 \xmm/EPIC-pn and \nus\ observations; cyan points indicate the 2006 \xmm/EPIC-pn spectrum; fuchsia, violet and orange points show the \chandra\ spectra in Obs.ID: 2026, 2027 and 2686, respectively. Spectra have been rebinned for display purposes only.}  
        \label{specallulx1}
\end{figure}

Finally, we analyzed the source spectra of the previous \xmm\ and \chandra\ observations. As a first step, we unfolded the spectra with a power-law of photon index 0 and we compared them with the 2019 {\it persistent} and {\it flare} spectra (Fig.~\ref{specallulx1}). We found that the 2003 \xmm\ and \chandra\ spectra are all compatible, with only subtle changes at soft energies, and in general consistent at energies above $\sim$5 keV with the 2019 {\it persistent} spectrum. 

We fitted the 2013 \xmm\ and \chandra\ spectra simultaneously, adopting an absorbed {\sc diskbb+diskPbb} model and a common absorption column. The fit gave a quite reasonable description of the data ($\chi^2_{\nu} = 492.3/465$), with a mean 0.3--10 keV absorbed flux of $\sim8\times10^{-13}$ erg cm$^{-2}$ s$^{-1}$, corresponding to a luminosity of $9.5\times10^{39}$ \lum. Results are reported in Table~\ref{spec_all}.




\section{Discussion}

We present the first long-term light curve of X7, by using all the available \swift, \xmm, \chandra\ and \nus\ observations taken between 2001 and 2020. Thanks to new high quality \xmm\ and \nus\ observations taken in 2019, we could investigate the source short-term temporal and spectral variability, and we characterized them in great detail.

\subsection{Flaring activity}

The source long-term flux evolution shows that X7 appears to be persistent (i.e. it is detected in each observation) and exhibits a flux variability of up to a factor of $\sim 6$ (from a luminosity of $\sim$$7\times10^{39}$ erg s$^{-1}$ to $\sim$$4\times10^{40}$ \lum, assuming a reference distance of 10 Mpc) over about 20 years of observations. We also report on a possible (super-)orbital variability with period of $\sim190$ d, to be confirmed by new continuous monitoring of the source. We note that super-orbital variabilities are seen in several ULXs with periods of tens to a few hundreds days \citep[e.g.][]{walton16,weng18, fuerst18,vasilopoulos20}, therefore the X7 periodicity would be in line with those. In addition, we remark that the possible X7 periodicity is better seen at high energies, for which disc precession effect may be stronger.

On ks time-scales, X7 is even more variable and shows clear flaring activity, which we observed for the first time in our most recent observations (2019). The flaring activity manifests itself only when the source is at its highest observed luminosities. At the peak of the flares, the luminosity was a factor of $\approx$3 higher than the pre-flare luminosity, which implies that the maximum variability can span almost an order of magnitude. About six flares were identified in the 2019 \xmm\ and \nus\ observations, with a typical duration of $\sim5 \text{ ks}-10 \text{ ks}$, and no obvious periodic or quasi-periodic pattern. Based on our data, flares appear to happen only at high fluxes (or luminosities higher than $2.5\times10^{40}$ \lum), and are quite rare. We could find flares only in 2 observations over a total of more than 50: no clear flares were observed in the \swift, \chandra\ and 2003 \xmm\ observations; although the number of \swift\ observations is large, we note that their exposure times are typically $\sim2$ ks, therefore shorter than the common flare duration, which prevents or strongly limits the detection of possible flares. Indeed, the scatter in flux of X7 in the \swift\ observations is quite wide (a factor of $\sim6$) and we cannot exclude that random bits of the flares are actually seen during the highest \swift\ luminosities $(\sim(2-3)\times 10^{40} \text{\lum})$. In particular, during the \swift\ observation 00032249048, we found that the source experienced a flux variability up to a factor of 3 suggesting a possible, partially seen, flare.

Furthermore, a prolonged high flux phase, which lasted about 15~ks, was identified at the end of the 2019 \xmm\ observation, and whose nature is unclear. During this time interval, the average flux was consistent with that observed at the peak of the flares, and shallow dip-like features can be observed. However, no significant spectral variations occur during the ``dips'', which indicates that most likely such features are accretion-driven, rather than e.g. induced by temporal increases in the local absorption, unless it is highly ionised (which would indeed be achromatic). It is also possible that a series of unresolved flares mimics the presence of dips. This possibility might be supported by the similar flux values reached during the isolated flares, and the high-flux phase. This behavior is reminiscent of one of the many variability classes of the BH binary GRS 1915+105 -- class $\kappa$ \citep[][]{belloni00} -- which is characterised by isolated events, followed by trains of flares with short recurrence times. Such a behavior, if observed in a significantly fainter source (which means that data quality will have a low count statistics), would be comparable with what we observed in the latest \xmm\ observation of X7.

Flaring activity is not common in ULXs and it was observed in the sources M51 ULX-7 \citep{earnshaw16a}, NGC 253 X-1 \citep{barnard10}, NGC 6946 ULX-3 \citep{earnshaw19b}, NGC 1313 X-1 \citep{walton20} and NGC 7456 ULX1 \citep{pintore20}. In none of them a (quasi-)periodicity of the flare recurrency was identified, as opposed to the quasi-periodic heart-beat of the ULX 4XMM J111816.0-324910 in the galaxy NGC 3621 \citep{motta20}. This may indicate that the flares in ULXs cannot be explained with only one mechanisms and, furthermore, whatever process that generates flares might or not generate periodic events. In fact, generally speaking the presence of flares does not appear to be clearly related to a specific spectral state or luminosity level (the range spanned by these source is very broad), although it seems that many ULXs preferentially flare during their brightest states. If we consider the spectral state of a given source, we remark they are quite different, which suggests that the high short-term activity can appear in both {\it soft ultraluminous}, {\it broadened disc} and {\it hard ultraluminous} states (where we adopted the spectral classification of \citealt{sutton13}). 

\subsection{Spectral evolution}
The spectral properties of X7 varied over the years, passing through a {\it soft-ultraluminous} state, with luminosities $<3\times 10^{40}$ \lum, to a {\it hard-ultraluminous} state characterised by flares. 
X7 is one of the best example of a ULX switching across two extreme ULX states. The general behaviour of the ULX population shows that sources can transit from {\it soft-ultraluminous} to {\it broadened disc} states, or from {\it hard-ultraluminous} to {\it broadened disc} states, or viceversa, but it is not often observed the transition from a {\it soft-ultraluminous} to a {\it hard-ultraluminous} state \citep[see, for example, the hardness-intensity/color-color diagrams in e.g.][]{sutton13, pintore17}. We have shown that the X7 spectra can be modelled with the combination of at least two thermal components, i.e. a standard multicolour blackbody disc ({\sc diskbb}) at low energies and a ``modified'' one ({\sc diskPbb}) at high energies. Such phenomenological model was proven to be a good representation of the spectra of a large sample of ULXs containing both PULXs and sources with compact objects of unknown nature \citep[e.g.][]{stobbart06, pintore15b, koliopanos17, walton20}. Based on our results it appears that, in X7, the behavior of the two disk components correlates with the luminosity (see Fig.~\ref{spec_flare_no_flare}). 

In particular, {\it at low luminosity}, the soft {\sc diskbb} component, which describes a standard Shakura-Sunyaev disk \citep{shakura73}, shows a temperature of $\sim$0.2 keV and an inner radius of $4.5^{+2.2}_{-1.3} \times 10^{3} \sqrt{(1/cos(i)}$ km  (where $i$ is the inclination angle of the system). The hard disk component, i.e. the {\sc diskPbb}, is likely to be ascribed to an advection dominated accretion flow (as suggested by the $\it p$ parameter, which assumes a value of $\sim$0.5), and its apparent temperature converged at $\sim$2 keV. Such a component is responsible for about the 70$\%$ of the 0.3--10 keV emission, therefore dominates the X-ray flux in this band. We constrained its apparent inner disc radius at $25^{+19}_{-10} \times \sqrt{(1/cos(i)}$ km. On the one hand, the small size of the radius of the hard disc component rules-out the presence of a disc around an IMBH. We stress that, should this radius be the innermost stable circular orbit ($\text{R}_{\text{isco}} = 6~GM/c^2$, for a non-rotating compact object), the mass of the compact object would be $\sim1.4-4$ M$_{\odot}$ and hence compatible with either a NS or a light BH. We note that for both disc components, the radii are estimated without taking into account any colour correction factor or boundary conditions: if we adopt a colour correction factor of $f_{col}=1.7$ (e.g. \citealt{shimura95}) and a boundary condition factor $\xi=0.4$ (e.g. \citealt{kubota98}), the radius estimates can be obtained from $R_{\text{in}} = \sqrt{K~(\xi~D~f_{col}^2)^2 / cos(i)}$ (where $K$ is the disc normalization, $D$ is the source distance): for the disc normalization values found in our spectral analysis, we calculated $R_{in_{\text{\sc diskbb}}} \sim 3000$ km and $R_{in_{\text{\sc diskPbb}}} \sim 16$ km, further strengthen that the compact object is likely of stellar origin (although its mass cannot be constrained yet in an obvious way).  

{\it At higher luminosity}, in epochs of the light-curve far from the flares, the spectra present an increase of the flux for energies below $\sim5$ keV, while at higher energies the source flux did not change significantly with respect to the lower flux spectra. A similar behaviour was identified in NGC 1313 X-1 and Holmberg IX X-1 \citep{walton17, walton20, gurpide21}, which are both relatively hard sources and both present a high-energy tail that dominates the spectrum above 10 keV. The origin of such a stable high energy tail has been explained in a super-Eddington accretion scenario, where the matter in the accretion flow is removed in an disk-fed outflow and this dampens most of the accretion rate variability at large radii, which leaves only low variability in the inner regions \citep{middleton15a} that therefore appear more stable. At super-Eddington rates, the outflows  are expected to form a funnel, which is responsible for the geometric beaming of the high energy emission. Such a scenario is one of the most likely explanation of the properties and behavior of the Galactic X-ray binary SS 433, which is believed to be an edge-on Galactic ULX \citep[][]{begelman06}. 
When the accretion rate increases further, the outflow opening angle can narrow \citep[e.g.][]{king09} even further, thus exacerbating the geometrical beaming of the high energy emission. Hence, in the sources with a hard-tail, the stability of the high energy component may indicate that, even though the flux is rising (i.e. the accretion rate is increasing), the emission of the wind is increasing as well, but the changes in accretion rate are not prominent enough to significantly affect the geometrical beaming. Alternatively, the increase of the accretion rate does not affect the aperture of the cone of the outflows, but rather it only increases the size of the apparent outflow emitting radius, i.e. the smallest disc radius at which outflows are ejected. In this last case the beaming may present a radial dependency, which implies that only the emission produced within a characteristic radius can be beamed \citep[see e.g.][]{lasota16, walton20}. In the case of X7, at least the first interpretation we described (i.e. a not changing geometrical beaming) may hold as the flux and spectral variability is mainly observed at low-middle energies, where the contribution of the outflow emission is thought to be dominant.

Instead, during the flares and at the highest source fluxes, X7 undergoes an abrupt spectral change with respect to the pre-flare state (see Fig.~\ref{lculx1}). While the flare and pre-flare spectra are consistent at energies below 1 keV, a clear increase in flux is observed above 1~keV during the flares. In addition, a third non-thermal spectral components may pop-up at energies $>10$ keV. Such a component is being detected in an increasing number of sources when high-quality and broadband data are available, and it is generally well-modelled by a cut-off power-law. Remarkably, this component tends to be stronger in PULXs (see e.g. \citealt{walton18}) and for this reason it is generally interpreted as the presence of the accretion column on the top of the NS in these sources. Our spectral modeling during the highest flux epochs and during the flares shows that the softer thermal component does not vary appreciably in flux with respect to the pre-flare epochs. This has a temperature of $\sim$0.2 keV and its corresponding emitting radius is $3.6^{+0.8}_{-0.6}\times 10^{3} \times \sqrt{1/cos(i)}$ km, consistent with the value derived for same component when observed at lower flux epochs. In contrast, the hard disc component shows an apparent inner radius of ($80^{+40}_{-20} \times \sqrt{1/cos(i)}$ km), which is larger than what we found at lower fluxes, although its temperature does not change significantly. 

\begin{figure*}
\center
\includegraphics[width=6.2cm, angle=90]{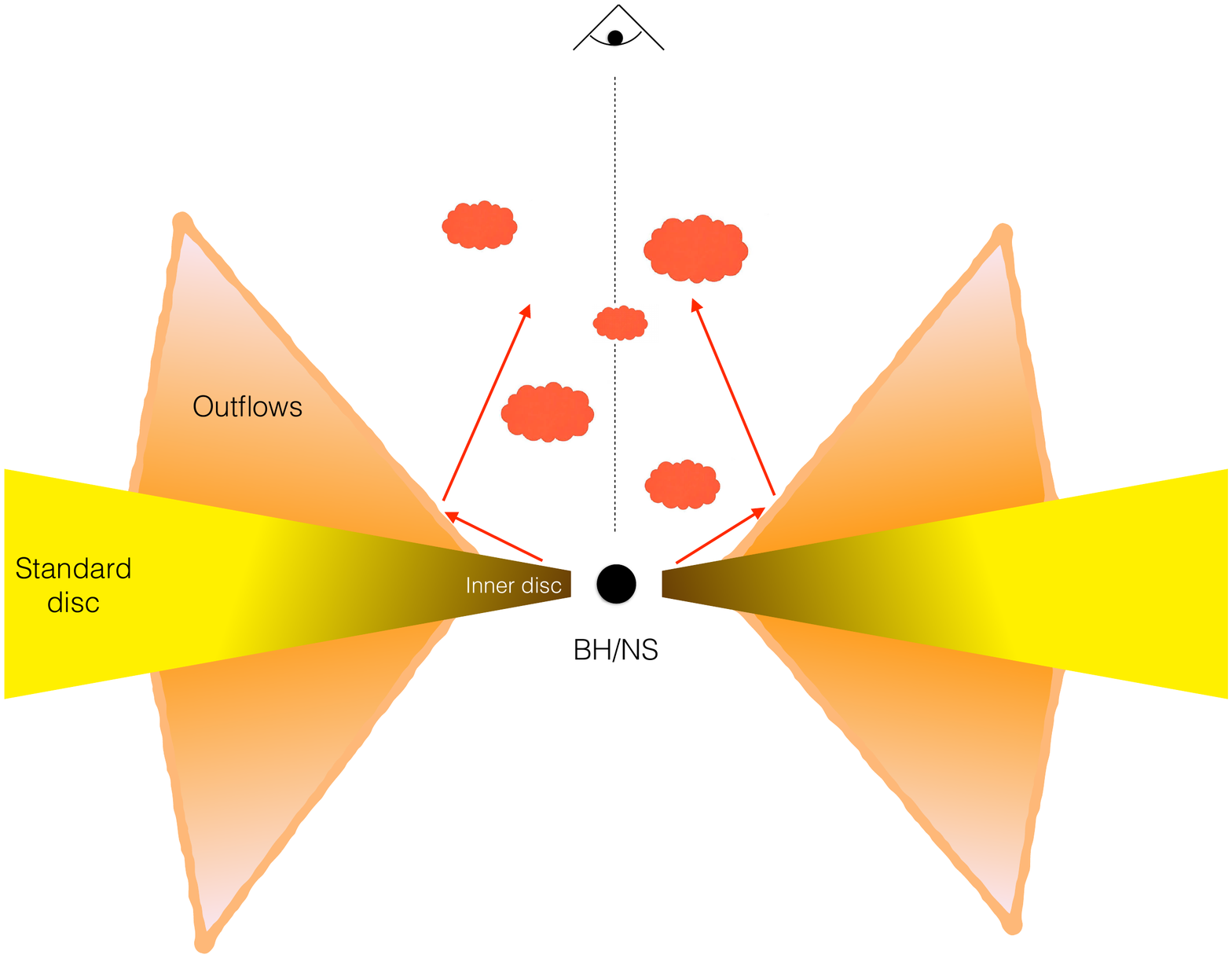}
\includegraphics[width=6.2cm, angle=90]{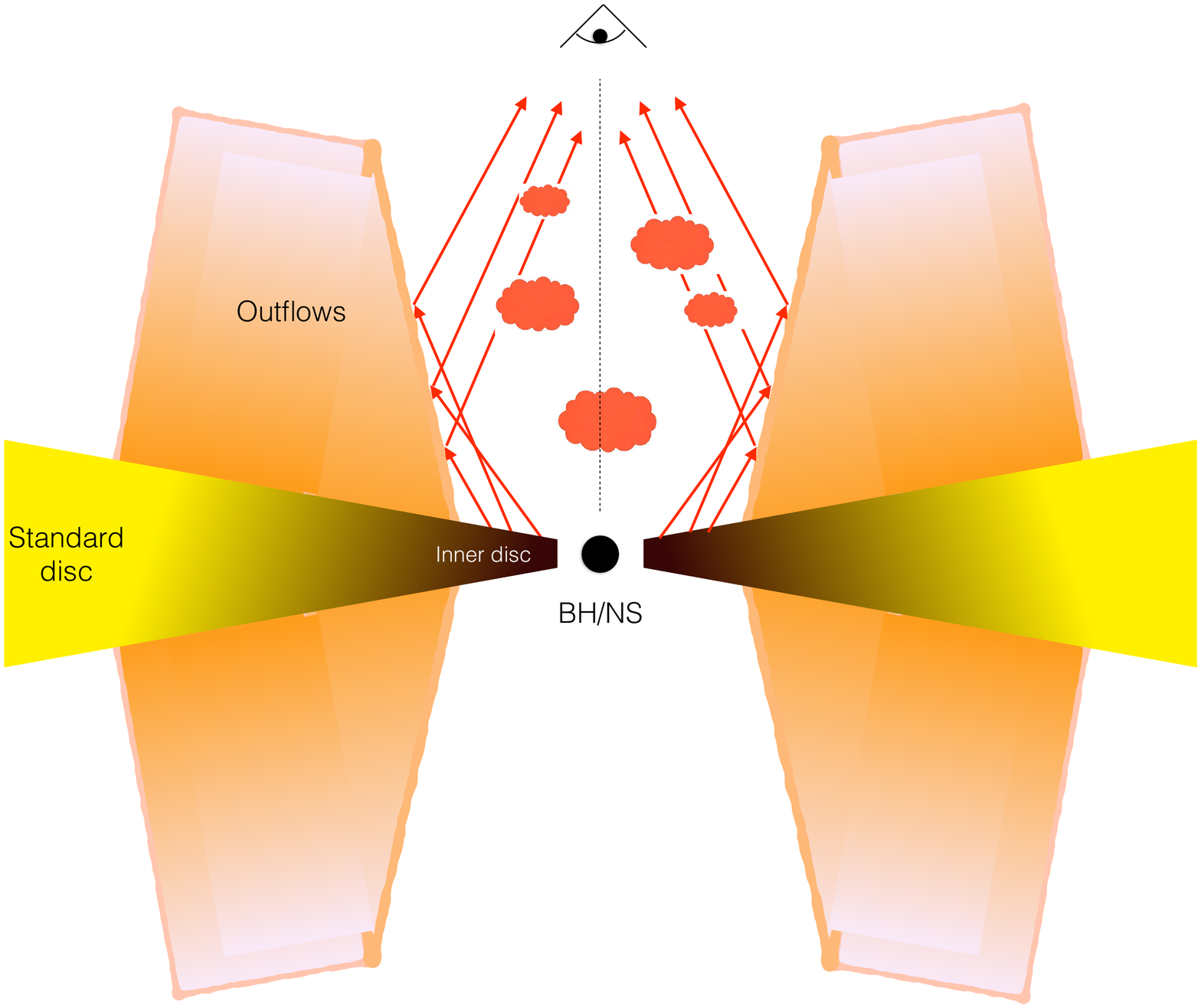}
   \caption{Schematic representation of the accretion flow in X7, where we assume that the accretion is super-Eddington and the LoS is close to be face-on. {\it Left}: The sketch shows the low flux state, in which a geometrical beaming of the inner regions is produced by the wall of the outflows. A standard accretion disc is expected in the outer regions. Optically thin turbulences are inside the wind cone producing the observed absorption and emission features. {\it Right}: the accretion flow during the flares and the high flux epochs. An enhancement of the accretion rate returns a narrowing of the wind cone opening angle, increasing the beaming of the inner regions.}  
        \label{sketch}
\end{figure*}

Based on our finding, and on the comparison between our results and what has been observed in other ULXs, we argue that each spectral component can be interpreted in the context of a super-Eddington accretion scenario, where the accretion flow is formed by: an outer disc; an advection dominated disc (or thick disc), which is covered by an optically thick outflow; and an inner accretion flow. The inner accretion flow could be a ``naked'' (i.e. without outflows) advection dominated disc or an optically thick corona covering the inner disc regions and the compact object. Alternatively it may be an optically thick boundary layer, should the compact object be a NS \citep[see]{mushtukov15}. We associate i) the outer disk to the softer thermal component ({\sc diskbb}), which we expect to be in a standard regime \citep[e.g.][]{shakura73}; ii) the thick-disc/outflows to the high-energy {\sc diskPbb} component; and iii) the cut-off power-law to the inner disc or the boundary layer.

\subsection{Evidences of outflows}
According to our spectral analysis, the soft disc component does not vary significantly in neither flux or temperature, which suggests that the size of its inner radius was constant even though the total flux increased by about an order of magnitude.
Because the data points are limited, we cannot claim any relation between the disc luminosity and its temperature. More high quality observations at different fluxes will allow us to investigate if the outer disc properties can be firmly associated to a standard disc. Similar considerations can be done for the high-energy disc component, which we associated to the photosphere of outflows ejected by the accretion disc. An analysis of the high-quality 2019 \xmm/RGS spectra allowed us to detect blue-shifted absorption and emission features. The former are probably due to O VII-VIII, while the latter are likely associated to Ne X or Fe L transitions. The significance of such spectral features is limited by the short exposure we had to employ, but the strict similarity with features reported in previous work (see, e.g., \citealt{pinto16,pinto17,kosec18b,wang19}) provides support for  the presence of optically-thin plasma and, likely, optically thick winds in X7. Given the likely presence of the winds, we note that the {\sc diskPbb} is a phenomenological model that can only mimic the more complex spectral emission of the outflows. We found that the flux evolution of this component is consistent with the photosphere of a wind increasing in size (the apparent emitting radius evolved from $\sim25$ km to $\sim80$ km) and decreasing in temperature because of the expansion. Our results indicate that the existence of outflows in X7 strengthens even more the compact object stellar origin, which we claimed to be likely a NS or a stellar-mass BH. Indeed, should the compact object being an IMBH, for the observed luminosity we would expect sub-Eddington or Eddington accretion rates, for which we do not expect powerful optically thick winds, nor an emission that spectrally bends below 10 keV.

Our results seem to indicate that X7 is seen from a small inclination angle, i.e. our line of sight intersects the cone produced by the wall of the optically thick outflow. In this scenario, the flares originate from a large increment of the accretion rate, which is able to significantly modify the geometrical structure of the outflows, in particular closing the aperture angle of the wind cone,  favouring a stronger geometrical beaming of the high-energy inner emission of the accretion flow. This results in a larger high energy flux, and thus in an hardening of the spectra  during the flaring activity (see the sketch in Fig.~\ref{sketch}).

\subsection{Short-term temporal properties}
The lags discovered in X7 can be explained in the super-Eddington framework as well. The hard lag is such that the high energy emission lags the soft one by $\sim300$s, on timescales of $\sim10^{-4}$ Hz. Hard lags are observed at low frequencies in a variety of Galactic X-ray binaries \citep{miyamoto88,arevalo06} and active galactic nuclei (AGN; e.g. \citealt{demarco13,alston13, uttley14, kara14, kara16}). Hard lags are commonly believed to be caused by the inward propagation of fluctuations in the accretion flow, for which the outer regions of the flow (responsible for the soft photons' emission) respond earlier than the inner regions where high-energy photons are produced \citep{kotov01, arevalo06}. X7 is the first ULX where low-frequency hard lags have been observed, with an average value of $334\pm89$s. We can explain such lags in terms of the inward propagation of mass accretion rate fluctuations which is also responsible for the flares. 

An alternative scenario may instead relate the low-frequency range of the lag to the temporal length between the flares: in such a case, a scattering of soft photons from a very extended optically thin corona or a jet may be likely. The corona can be supported by the RGS spectroscopic analysis, which showed the existence of emission possibly produced in an optically thin plasma. Should the hard lag simply be the crossing time of the soft photons into a hot and optically thin medium, it implies a size of $\sim10^{13}$ cm, which may be too large and unfeasible. Hence, the inward propagating fluctuations remain the best explanation for the source hard lag.

In addition, we also remark the existence of a weak soft lag at frequencies $\sim(2-4)\times10^{-4}$ Hz with a weighted average value of $395 \pm 167$ s. Soft lags were reported for the sources NGC 55 ULX-1 \citep{pinto17}, NGC 1313 X-1 \citep{kara20} and NGC 5408 X-1 \citep{demarco13}, with magnitudes from tens to thousands of ks. Soft lags are commonly associated with delays due to the crossing time and down-scattering of hard photons passing through the optically thick outflows. The interpretation of the soft lags in these sources then points towards a quite high inclination angle of the systems. However, the phenomenology of X7 seems instead to point towards a quite low inclination of the system. Should the soft lag in X7 be real, this would be in line with the magnitudes observed in the other ULXs, but it may be interpreted as reflection of the hard photons from a lower temperature region. Assuming that the Compton scattering is the dominant process, the radius of the scattering region can be estimated as $R=\sqrt{\cfrac{c~t }{\sigma_T~n_e}}$, where $c$ is the speed of light, $t$ is the soft lag in seconds, $\sigma_T$ is the Thompson cross-scattering and $n_e$ is the electron density. Assuming a disc/wind electron density of $n_e\sim10^{20}$ cm$^{-3}$ \citep[e.g.][]{shakura73, takeuchi13} and a single scattering, the size of the scattering medium would then be $<10^7-10^8$ cm, i.e. larger than the characteristic inner disc radii estimated from the spectral analysis, and instead consistent with the  outer regions of the accretion flow or with the base of the outflow. We note that the estimated size could be smaller if we allow for multi-scatterings of the hard photons into the medium.  

Finally, although we analyzed the highest quality available data of the source, which shed light on the geometry and structure of the accretion flow, the exact nature of the compact object in X7 remains dubious. The timing analysis did not provide any strong indication on the existence of pulsations, but only some hints of pulsations when orbital corrections are applied, the statistical significance of which is very low. More high quality data are needed to confirm or refute such a tentative pulsation detection. The large flux variation on long timescale, the hard spectra during the flares and the small size of the inner disc during the low flux epochs can favour X7 hosting a NS. We note that, should the compact object be a magnetic NS able to truncate the inner accretion flow at radii of tens of km, as suggested by our spectral analysis, by assuming a beaming factor of 0.1, we derive a dipolar magnetic field of magnitude $10^{10}-10^{11}$ G.

\section{Conclusions}

We analyzed all the available X-ray observations of the ULX  X7 taken with \xmm, \nus, \chandra\ and \swift. Its long-term variability features changes in flux of up to an order of magnitude, with extreme luminosity reaching an observed maximum of $\sim6\times10^{40}$ \lum\ during flare activity. The spectral analysis of the high quality \xmm\ and \nus\ observations allowed us to characterize the properties of the source, which we modelled with two thermal disc-like components. We interpret the source spectral and temporal evolution within the framework of super-Eddington accretion and we associated the soft disc component to an outer, standard razor-thin accretion disc, while the high-energy disc to the photosphere of a powerful optically thick outflow above the disc. Confirmation on the existence of a wind comes from the detection of blue-shifted absorption and emission features in the RGS data, which indicate also the presence of an optically thin plasma. We also note that during the flare, a third spectral component could be exhibited by X7. Comparing it with the ones observed in a number of ULXs, we may associate it to either an accretion column density if the compact object is a NS or a boundary layer/optically thick corona in case of a BH accretor. 
We propose that the source evolution can be explained with variations of the accretion rate that can affect the geometrical cone of the outflow, enhancing the degree of beaming of the inner regions responsible for the high-energy emission.

\section*{Acknowledgements} 

This work has been partially supported by the ASI-INAF agreements 2017-14-H.0.

\section*{Data Availability}
All of the data underlying this article are already publicly available from ESA's XMM-Newton Science Archive (https://www.cosmos.esa.int/web/xmm-newton/xsa), NASA's HEASARC archive (https://heasarc.gsfc.nasa.gov/), and the Chandra Data Archive (https://cxc.harvard.edu/cda/).

\addcontentsline{toc}{section}{Bibliography}
\bibliographystyle{mn2e}
\bibliography{biblio}

\begin{thebibliography}{}

\bibitem[\protect\citeauthoryear{{Alston}, {Vaughan} \& {Uttley}}{{Alston}
  et~al.}{2013}]{alston13}
{Alston} W.~N.,  {Vaughan} S.,    {Uttley} P.,  2013, \mnras, 435, 1511

\bibitem[\protect\citeauthoryear{{Ar{\'e}valo} \& {Uttley}}{{Ar{\'e}valo} \&
  {Uttley}}{2006}]{arevalo06}
{Ar{\'e}valo} P.,  {Uttley} P.,  2006, \mnras, 367, 801

\bibitem[\protect\citeauthoryear{{Bachetti}, {Harrison}, {Walton},
  {Grefenstette}, {Chakrabarty}, {F{\"u}rst}, {Barret} \& et al.}{{Bachetti}
  et~al.}{2014}]{bachetti14}
{Bachetti} M.,  {Harrison} F.~A.,  {Walton} D.~J.,  {Grefenstette} B.~W.,
  {Chakrabarty} D.,  {F{\"u}rst} F.,  {Barret}   et al. 2014, \nat, 514, 202

\bibitem[\protect\citeauthoryear{{Bachetti}, {Rana}, {Walton}, {Barret},
  {Harrison} \& et al.}{{Bachetti} et~al.}{2013}]{bachetti13}
{Bachetti} M.,  {Rana} V.,  {Walton} D.~J.,  {Barret} D.,  {Harrison} F.~A.,
  et al. 2013, \apj, 778, 163

\bibitem[\protect\citeauthoryear{{Band}}{{Band}}{1997}]{Band97}
{Band} D.~L.,  1997, \apj, 486, 928

\bibitem[\protect\citeauthoryear{{Barnard}}{{Barnard}}{2010}]{barnard10}
{Barnard} R.,  2010, \mnras, 404, 42

\bibitem[\protect\citeauthoryear{{Begelman}, {King} \& {Pringle}}{{Begelman}
  et~al.}{2006}]{begelman06}
{Begelman} M.~C.,  {King} A.~R.,    {Pringle} J.~E.,  2006, \mnras, 370, 399

\bibitem[\protect\citeauthoryear{{Belloni}, {Klein-Wolt}, {M{\'e}ndez}, {van
  der Klis} \& {van Paradijs}}{{Belloni} et~al.}{2000}]{belloni00}
{Belloni} T.,  {Klein-Wolt} M.,  {M{\'e}ndez} M.,  {van der Klis} M.,    {van
  Paradijs} J.,  2000, \aap, 355, 271

\bibitem[\protect\citeauthoryear{{Brightman}, {Harrison}, {F{\"u}rst},
  {Middleton}, {Walton}, {Stern}, {Fabian}, {Heida}, {Barret} \&
  {Bachetti}}{{Brightman} et~al.}{2018}]{brightman18}
{Brightman} M.,  {Harrison} F.~A.,  {F{\"u}rst} F.,  {Middleton} M.~J.,
  {Walton} D.~J.,  {Stern} D.,  {Fabian} A.~C.,  {Heida} M.,  {Barret} D.,
  {Bachetti} M.,  2018, Nature Astronomy, 2, 312

\bibitem[\protect\citeauthoryear{{Carpano}, {Haberl}, {Maitra} \&
  {Vasilopoulos}}{{Carpano} et~al.}{2018}]{carpano18}
{Carpano} S.,  {Haberl} F.,  {Maitra} C.,    {Vasilopoulos} G.,  2018, \mnras

\bibitem[\protect\citeauthoryear{{Cash}}{{Cash}}{1979}]{Cash1979}
{Cash} W.,  1979, \apj, 228, 939

\bibitem[\protect\citeauthoryear{{De Marco}, {Ponti}, {Miniutti}, {Belloni},
  {Cappi}, {Dadina} \& {Mu{\~n}oz-Darias}}{{De Marco} et~al.}{2013}]{demarco13}
{De Marco} B.,  {Ponti} G.,  {Miniutti} G.,  {Belloni} T.,  {Cappi} M.,
  {Dadina} M.,    {Mu{\~n}oz-Darias} T.,  2013, \mnras, 436, 3782

\bibitem[\protect\citeauthoryear{{den Herder}, {Brinkman}, {Kahn} et~al.,}{{den
  Herder} et~al.}{2001}]{denHerder2001}
{den Herder} J.~W.,  {Brinkman} A.~C.,  {Kahn} S.~M.,    et~al., 2001, A\&A,
  365, L7

\bibitem[\protect\citeauthoryear{{Earnshaw}, {Roberts}, {Heil}, {Mezcua},
  {Walton}, {Done}, {Harrison}, {Lansbury}, {Middleton} \& {Sutton}}{{Earnshaw}
  et~al.}{2016}]{earnshaw16a}
{Earnshaw} H.~M.,  {Roberts} T.~P.,  {Heil} L.~M.,  {Mezcua} M.,  {Walton}
  D.~J.,  {Done} C.,  {Harrison} F.~A.,  {Lansbury} G.~B.,  {Middleton} M.~J.,
    {Sutton} A.~D.,  2016, \mnras, 456, 3840

\bibitem[\protect\citeauthoryear{{Earnshaw}, {Grefenstette}, {Brightman},
  {Walton}, {Barret}, {F{\"u}rst}, {Harrison}, {Heida}, {Pike}, {Stern} \&
  {Webb}}{{Earnshaw} et~al.}{2019}]{earnshaw19b}
{Earnshaw} H.~P.,  {Grefenstette} B.~W.,  {Brightman} M.,  {Walton} D.~J.,
  {Barret} D.,  {F{\"u}rst} F.,  {Harrison} F.~A.,  {Heida} M.,  {Pike} S.~N.,
  {Stern} D.,    {Webb} N.~A.,  2019, \apj, 881, 38

\bibitem[\protect\citeauthoryear{{Earnshaw}, {Roberts} \&
  {Sathyaprakash}}{{Earnshaw} et~al.}{2018}]{earnshaw18}
{Earnshaw} H.~P.,  {Roberts} T.~P.,    {Sathyaprakash} R.,  2018, \mnras, 476,
  4272

\bibitem[\protect\citeauthoryear{{F{\"u}rst}, {Walton}, {Heida}, {Harrison},
  {Barret}, {Brightman}, {Fabian}, {Middleton}, {Pinto}, {Rana}, {Tramper},
  {Webb} \& {Kretschmar}}{{F{\"u}rst} et~al.}{2018}]{fuerst18}
{F{\"u}rst} F.,  {Walton} D.~J.,  {Heida} M.,  {Harrison} F.~A.,  {Barret} D.,
  {Brightman} M.,  {Fabian} A.~C.,  {Middleton} M.~J.,  {Pinto} C.,  {Rana} V.,
   {Tramper} F.,  {Webb} N.,    {Kretschmar} P.,  2018, \aap, 616, A186

\bibitem[\protect\citeauthoryear{{G{\'u}rpide}, {Godet}, {Koliopanos}, {Webb}
  \& {Olive}}{{G{\'u}rpide} et~al.}{2021}]{gurpide21}
{G{\'u}rpide} A.,  {Godet} O.,  {Koliopanos} F.,  {Webb} N.,    {Olive} J.-F.,
  2021, arXiv e-prints, p. arXiv:2102.11159

\bibitem[\protect\citeauthoryear{{Heil} \& {Vaughan}}{{Heil} \&
  {Vaughan}}{2010}]{heil10}
{Heil} L.~M.,  {Vaughan} S.,  2010, \mnras, 405, L86

\bibitem[\protect\citeauthoryear{{Israel}, {Belfiore}, {Stella} \&
  {Esposito}}{{Israel} et~al.}{2017}]{israel16a}
{Israel} G.~L.,  {Belfiore} A.,  {Stella} L.,    {Esposito} P. e.~a.,  2017,
  Science, 355, 817

\bibitem[\protect\citeauthoryear{{Israel}, {Papitto}, {Esposito} \&
  {Stella}}{{Israel} et~al.}{2017}]{israel16b}
{Israel} G.~L.,  {Papitto} A.,  {Esposito} P.,    {Stella} L. e.~a.,  2017,
  \mnras, 466, L48

\bibitem[\protect\citeauthoryear{{Israel} \& {Stella}}{{Israel} \&
  {Stella}}{1996}]{israel96}
{Israel} G.~L.,  {Stella} L.,  1996, \apj, 468, 369

\bibitem[\protect\citeauthoryear{{Kaaret}, {Feng} \& {Roberts}}{{Kaaret}
  et~al.}{2017}]{kaaret17}
{Kaaret} P.,  {Feng} H.,    {Roberts} T.~P.,  2017, \araa, 55, 303

\bibitem[\protect\citeauthoryear{{Kaastra}, {Mewe} \&
  {Nieuwenhuijzen}}{{Kaastra} et~al.}{1996}]{Kaastra1996}
{Kaastra} J.~S.,  {Mewe} R.,    {Nieuwenhuijzen} H.,  1996, in {K.~Yamashita \&
  T.~Watanabe} ed., UV and X-ray Spectroscopy of Astrophysical and Laboratory
  Plasmas {SPEX: a new code for spectral analysis of X {\&} UV spectra.}.
p.~411

\bibitem[\protect\citeauthoryear{{Kara}, {Alston}, {Fabian}, {Cackett},
  {Uttley}, {Reynolds} \& {Zoghbi}}{{Kara} et~al.}{2016}]{kara16}
{Kara} E.,  {Alston} W.~N.,  {Fabian} A.~C.,  {Cackett} E.~M.,  {Uttley} P.,
  {Reynolds} C.~S.,    {Zoghbi} A.,  2016, \mnras, 462, 511

\bibitem[\protect\citeauthoryear{{Kara}, {Fabian}, {Marinucci}, {Matt},
  {Parker}, {Alston}, {Brenneman}, {Cackett} \& {Miniutti}}{{Kara}
  et~al.}{2014}]{kara14}
{Kara} E.,  {Fabian} A.~C.,  {Marinucci} A.,  {Matt} G.,  {Parker} M.~L.,
  {Alston} W.,  {Brenneman} L.~W.,  {Cackett} E.~M.,    {Miniutti} G.,  2014,
  \mnras, 445, 56

\bibitem[\protect\citeauthoryear{{Kara}, {Pinto}, {Walton}, {Alston},
  {Bachetti}, {Barret}, {Brightman}, {Canizares}, {Earnshaw}, {Fabian},
  {F{\"u}rst}, {Kosec}, {Middleton}, {Roberts}, {Soria}, {Tao} \&
  {Webb}}{{Kara} et~al.}{2020}]{kara20}
{Kara} E.,  {Pinto} C.,  {Walton} D.~J.,  {Alston} W.~N.,  {Bachetti} M.,
  {Barret} D.,  {Brightman} M.,  {Canizares} C.~R.,  {Earnshaw} H.~P.,
  {Fabian} A.~C.,  {F{\"u}rst} F.,  {Kosec} P.,  {Middleton} M.~J.,  {Roberts}
  T.~P.,  {Soria} R.,  {Tao} L.,    {Webb} N.~A.,  2020, \mnras, 491, 5172

\bibitem[\protect\citeauthoryear{{King} \& {Lasota}}{{King} \&
  {Lasota}}{2020}]{king20}
{King} A.,  {Lasota} J.-P.,  2020, \mnras, 494, 3611

\bibitem[\protect\citeauthoryear{{King}}{{King}}{2009}]{king09}
{King} A.~R.,  2009, \mnras, 393, L41

\bibitem[\protect\citeauthoryear{{Koliopanos}, {Vasilopoulos}, {Godet},
  {Bachetti}, {Webb} \& {Barret}}{{Koliopanos} et~al.}{2017}]{koliopanos17}
{Koliopanos} F.,  {Vasilopoulos} G.,  {Godet} O.,  {Bachetti} M.,  {Webb}
  N.~A.,    {Barret} D.,  2017, \aap, 608, A47

\bibitem[\protect\citeauthoryear{{Kosec}, {Pinto}, {Fabian} \&
  {Walton}}{{Kosec} et~al.}{2018}]{kosec18a}
{Kosec} P.,  {Pinto} C.,  {Fabian} A.~C.,    {Walton} D.~J.,  {2018}, \mnras,
  473, 5680

\bibitem[\protect\citeauthoryear{{Kosec}, {Pinto}, {Walton}, {Fabian},
  {Bachetti}, {Brightman}, {F{\"u}rst} \& {Grefenstette}}{{Kosec}
  et~al.}{2018}]{kosec18b}
{Kosec} P.,  {Pinto} C.,  {Walton} D.~J.,  {Fabian} A.~C.,  {Bachetti} M.,
  {Brightman} M.,  {F{\"u}rst} F.,    {Grefenstette} B.~W.,  2018, \mnras, 479,
  3978

\bibitem[\protect\citeauthoryear{{Kotov}, {Churazov} \& {Gilfanov}}{{Kotov}
  et~al.}{2001}]{kotov01}
{Kotov} O.,  {Churazov} E.,    {Gilfanov} M.,  2001, \mnras, 327, 799

\bibitem[\protect\citeauthoryear{{Kubota}, {Tanaka}, {Makishima}, {Ueda},
  {Dotani}, {Inoue} \& {Yamaoka}}{{Kubota} et~al.}{1998}]{kubota98}
{Kubota} A.,  {Tanaka} Y.,  {Makishima} K.,  {Ueda} Y.,  {Dotani} T.,  {Inoue}
  H.,    {Yamaoka} K.,  1998, \pasj, 50, 667

\bibitem[\protect\citeauthoryear{{Lasota}, {Vieira}, {Sadowski}, {Narayan} \&
  {Abramowicz}}{{Lasota} et~al.}{2016}]{lasota16}
{Lasota} J.~P.,  {Vieira} R.~S.~S.,  {Sadowski} A.,  {Narayan} R.,
  {Abramowicz} M.~A.,  2016, \aap, 587, A13

\bibitem[\protect\citeauthoryear{{Middleton}, {Brightman}, {Pintore},
  {Bachetti}, {Fabian}, {F{\"u}rst} \& {Walton}}{{Middleton}
  et~al.}{2019}]{middleton19}
{Middleton} M.~J.,  {Brightman} M.,  {Pintore} F.,  {Bachetti} M.,  {Fabian}
  A.~C.,  {F{\"u}rst} F.,    {Walton} D.~J.,  2019, \mnras, 486, 2

\bibitem[\protect\citeauthoryear{{Middleton}, {Heil}, {Pintore}, {Walton} \&
  {Roberts}}{{Middleton} et~al.}{2015a}]{middleton15a}
{Middleton} M.~J.,  {Heil} L.,  {Pintore} F.,  {Walton} D.~J.,    {Roberts}
  T.~P.,  2015a, \mnras, 447, 3243

\bibitem[\protect\citeauthoryear{{Middleton} \& {King}}{{Middleton} \&
  {King}}{2017}]{middleton17}
{Middleton} M.~J.,  {King} A.,  2017, \mnras, 470, L69

\bibitem[\protect\citeauthoryear{{Middleton}, {Walton}, {Fabian}, {Roberts},
  {Heil}, {Pinto}, {Anderson} \& {Sutton}}{{Middleton}
  et~al.}{2015b}]{middleton15b}
{Middleton} M.~J.,  {Walton} D.~J.,  {Fabian} A.,  {Roberts} T.~P.,  {Heil} L.,
   {Pinto} C.,  {Anderson} G.,    {Sutton} A.,  2015b, \mnras, 454, 3134

\bibitem[\protect\citeauthoryear{{Miyamoto}, {Kitamoto}, {Mitsuda} \&
  {Dotani}}{{Miyamoto} et~al.}{1988}]{miyamoto88}
{Miyamoto} S.,  {Kitamoto} S.,  {Mitsuda} K.,    {Dotani} T.,  1988, \nat, 336,
  450

\bibitem[\protect\citeauthoryear{{Motta}, {Marelli}, {Pintore}, {Esposito},
  {Salvaterra}, {De Luca}, {Israel}, {Tiengo} \& {Castillo}}{{Motta}
  et~al.}{2020}]{motta20}
{Motta} S.~E.,  {Marelli} M.,  {Pintore} F.,  {Esposito} P.,  {Salvaterra} R.,
  {De Luca} A.,  {Israel} G.~L.,  {Tiengo} A.,    {Castillo} G.~A.~R.,  2020,
  \apj, 898, 174

\bibitem[\protect\citeauthoryear{{Mushtukov}, {Suleimanov}, {Tsygankov} \&
  {Poutanen}}{{Mushtukov} et~al.}{2015}]{mushtukov15}
{Mushtukov} A.~A.,  {Suleimanov} V.~F.,  {Tsygankov} S.~S.,    {Poutanen} J.,
  2015, \mnras, 454, 2539

\bibitem[\protect\citeauthoryear{{Ohsuga} \& {Mineshige}}{{Ohsuga} \&
  {Mineshige}}{2011}]{ohsuga11}
{Ohsuga} K.,  {Mineshige} S.,  2011, \apj, 736, 2

\bibitem[\protect\citeauthoryear{{Peterson}, {Wanders}, {Horne}, {Collier},
  {Alexander}, {Kaspi} \& {Maoz}}{{Peterson} et~al.}{1998}]{Peterson98}
{Peterson} B.~M.,  {Wanders} I.,  {Horne} K.,  {Collier} S.,  {Alexander} T.,
  {Kaspi} S.,    {Maoz} D.,  1998, \pasp, 110, 660

\bibitem[\protect\citeauthoryear{{Pinto}, {Alston}, {Soria}, {Middleton},
  {Walton}, {Sutton}, {Fabian}, {Earnshaw}, {Urquhart}, {Kara} \&
  {Roberts}}{{Pinto} et~al.}{2017}]{pinto17}
{Pinto} C.,  {Alston} W.,  {Soria} R.,  {Middleton} M.~J.,  {Walton} D.~J.,
  {Sutton} A.~D.,  {Fabian} A.~C.,  {Earnshaw} H.,  {Urquhart} R.,  {Kara} E.,
    {Roberts} T.~P.,  2017, \mnras, 468, 2865

\bibitem[\protect\citeauthoryear{{Pinto}, {Middleton} \& {Fabian}}{{Pinto}
  et~al.}{2016}]{pinto16}
{Pinto} C.,  {Middleton} M.~J.,    {Fabian} A.~C.,  2016, \nat, 533, 64

\bibitem[\protect\citeauthoryear{{Pinto}, {Walton}, {Kara}, {Parker}, {Soria},
  {Kosec}, {Middleton}, {Alston}, {Fabian}, {Guainazzi}, {Roberts}, {Fuerst},
  {Earnshaw}, {Sathyaprakash} \& {Barret}}{{Pinto} et~al.}{2020}]{pinto20}
{Pinto} C.,  {Walton} D.~J.,  {Kara} E.,  {Parker} M.~L.,  {Soria} R.,  {Kosec}
  P.,  {Middleton} M.~J.,  {Alston} W.~N.,  {Fabian} A.~C.,  {Guainazzi} M.,
  {Roberts} T.~P.,  {Fuerst} F.,  {Earnshaw} H.~P.,  {Sathyaprakash} R.,
  {Barret} D.,  2020, \mnras, 492, 4646

\bibitem[\protect\citeauthoryear{{Pintore}, {Esposito}, {Zampieri}, {Motta} \&
  {Wolter}}{{Pintore} et~al.}{2015}]{pintore15b}
{Pintore} F.,  {Esposito} P.,  {Zampieri} L.,  {Motta} S.,    {Wolter} A.,
  2015, \mnras, 448, 1153

\bibitem[\protect\citeauthoryear{{Pintore}, {Marelli}, {Salvaterra}, {Israel}
  \& {et al.}}{{Pintore} et~al.}{2020}]{pintore20}
{Pintore} F.,  {Marelli} M.,  {Salvaterra} R.,  {Israel} G.~L.,    {et al.}
  2020, \apj, 890, 166

\bibitem[\protect\citeauthoryear{{Pintore}, {Zampieri}, {Stella}, {Wolter},
  {Mereghetti} \& {Israel}}{{Pintore} et~al.}{2017}]{pintore17}
{Pintore} F.,  {Zampieri} L.,  {Stella} L.,  {Wolter} A.,  {Mereghetti} S.,
  {Israel} G.~L.,  2017, \apj, 836, 113

\bibitem[\protect\citeauthoryear{{Poutanen}, {Lipunova}, {Fabrika}, {Butkevich}
  \& {Abolmasov}}{{Poutanen} et~al.}{2007}]{poutanen07}
{Poutanen} J.,  {Lipunova} G.,  {Fabrika} S.,  {Butkevich} A.~G.,
  {Abolmasov} P.,  2007, \mnras, 377, 1187

\bibitem[\protect\citeauthoryear{{Rodr{\'\i}guez Castillo}, {Israel},
  {Belfiore}, {Bernardini}, {Esposito}, {Pintore}, {De Luca}, {Papitto},
  {Stella} \& et al.}{{Rodr{\'\i}guez Castillo} et~al.}{2019}]{rodriguez19}
{Rodr{\'\i}guez Castillo} G.~A.,  {Israel} G.~L.,  {Belfiore} A.,  {Bernardini}
  F.,  {Esposito} P.,  {Pintore} F.,  {De Luca} A.,  {Papitto} A.,  {Stella}
  L.,    et al. 2019, \apj, submitted (eprint: astro-ph.HE/1906.04791)

\bibitem[\protect\citeauthoryear{{Sanders}, {Mazzarella}, {Kim}, {Surace} \&
  {Soifer}}{{Sanders} et~al.}{2003}]{sanders03}
{Sanders} D.~B.,  {Mazzarella} J.~M.,  {Kim} D.-C.,  {Surace} J.~A.,
  {Soifer} B.~T.,  2003, \aj, 126, 1607

\bibitem[\protect\citeauthoryear{{Sathyaprakash}, {Roberts}, {Walton},
  {Fuerst}, {Bachetti}, {Pinto}, {Alston}, {Earnshaw}, {Fabian} \&
  {Middleton}}{{Sathyaprakash} et~al.}{2019}]{sathyaprakash19}
{Sathyaprakash} R.,  {Roberts} T.~P.,  {Walton} D.~J.,  {Fuerst} F.,
  {Bachetti} M.,  {Pinto} C.,  {Alston} W.~N.,  {Earnshaw} H.~P.,  {Fabian}
  A.~C.,    {Middleton} M.~J.,  2019, arXiv e-prints, p. arXiv:1906.00640

\bibitem[\protect\citeauthoryear{{Shakura} \& {Sunyaev}}{{Shakura} \&
  {Sunyaev}}{1973}]{shakura73}
{Shakura} N.~I.,  {Sunyaev} R.~A.,  1973, \aap, 24, 337

\bibitem[\protect\citeauthoryear{{Shimura} \& {Takahara}}{{Shimura} \&
  {Takahara}}{1995}]{shimura95}
{Shimura} T.,  {Takahara} F.,  1995, \apj, 445, 780

\bibitem[\protect\citeauthoryear{{Sorce}, {Tully}, {Courtois}, {Jarrett},
  {Neill} \& {Shaya}}{{Sorce} et~al.}{2014}]{sorce14}
{Sorce} J.~G.,  {Tully} R.~B.,  {Courtois} H.~M.,  {Jarrett} T.~H.,  {Neill}
  J.~D.,    {Shaya} E.~J.,  2014, \mnras, 444, 527

\bibitem[\protect\citeauthoryear{{Soria}, {Motch}, {Read} \& {Stevens}}{{Soria}
  et~al.}{2004}]{soria04}
{Soria} R.,  {Motch} C.,  {Read} A.~M.,    {Stevens} I.~R.,  2004, \aap, 423,
  955

\bibitem[\protect\citeauthoryear{{Stobbart}, {Roberts} \& {Warwick}}{{Stobbart}
  et~al.}{2004}]{stobbart04}
{Stobbart} A.-M.,  {Roberts} T.~P.,    {Warwick} R.~S.,  2004, \mnras, 351,
  1063

\bibitem[\protect\citeauthoryear{{Stobbart}, {Roberts} \& {Wilms}}{{Stobbart}
  et~al.}{2006}]{stobbart06}
{Stobbart} A.-M.,  {Roberts} T.~P.,    {Wilms} J.,  2006, \mnras, 368, 397

\bibitem[\protect\citeauthoryear{{Sutton}, {Roberts} \& {Middleton}}{{Sutton}
  et~al.}{2013}]{sutton13}
{Sutton} A.~D.,  {Roberts} T.~P.,    {Middleton} M.~J.,  2013, \mnras, 435,
  1758

\bibitem[\protect\citeauthoryear{{Swartz}, {Soria}, {Tennant} \&
  {Yukita}}{{Swartz} et~al.}{2011}]{swartz11}
{Swartz} D.~A.,  {Soria} R.,  {Tennant} A.~F.,    {Yukita} M.,  2011, \apj,
  741, 49

\bibitem[\protect\citeauthoryear{{Takeuchi}, {Ohsuga} \&
  {Mineshige}}{{Takeuchi} et~al.}{2013}]{takeuchi13}
{Takeuchi} S.,  {Ohsuga} K.,    {Mineshige} S.,  2013, \pasj, 65, 88

\bibitem[\protect\citeauthoryear{{Takeuchi}, {Ohsuga} \&
  {Mineshige}}{{Takeuchi} et~al.}{2014}]{takeuchi14}
{Takeuchi} S.,  {Ohsuga} K.,    {Mineshige} S.,  2014, \pasj

\bibitem[\protect\citeauthoryear{{Tsygankov}, {Mushtukov}, {Suleimanov} \&
  {Poutanen}}{{Tsygankov} et~al.}{2016}]{tsygankov16}
{Tsygankov} S.~S.,  {Mushtukov} A.~A.,  {Suleimanov} V.~F.,    {Poutanen} J.,
  2016, \mnras, 457, 1101

\bibitem[\protect\citeauthoryear{{Tully} \& {Fisher}}{{Tully} \&
  {Fisher}}{1988}]{tully88}
{Tully} R.~B.,  {Fisher} J.~R.,  1988, {Catalog of Nearby Galaxies}

\bibitem[\protect\citeauthoryear{{Uttley}, {Cackett}, {Fabian}, {Kara} \&
  {Wilkins}}{{Uttley} et~al.}{2014}]{uttley14}
{Uttley} P.,  {Cackett} E.~M.,  {Fabian} A.~C.,  {Kara} E.,    {Wilkins} D.~R.,
   2014, \aapr, 22, 72

\bibitem[\protect\citeauthoryear{{Vasilopoulos}, {Lander}, {Koliopanos} \&
  {Bailyn}}{{Vasilopoulos} et~al.}{2020}]{vasilopoulos20}
{Vasilopoulos} G.,  {Lander} S.~K.,  {Koliopanos} F.,    {Bailyn} C.~D.,  2020,
  \mnras, 491, 4949

\bibitem[\protect\citeauthoryear{{Walton}, {F{\"u}rst}, {Bachetti}, {Barret},
  {Brightman}, {Fabian}, {Gehrels}, {Harrison}, {Heida}, {Middleton}, {Rana},
  {Roberts}, {Stern}, {Tao} \& {Webb}}{{Walton} et~al.}{2016a}]{walton16}
{Walton} D.~J.,  {F{\"u}rst} F.,  {Bachetti} M.,  {Barret} D.,  {Brightman} M.,
   {Fabian} A.~C.,  {Gehrels} N.,  {Harrison} F.~A.,  {Heida} M.,  {Middleton}
  M.~J.,  {Rana} V.,  {Roberts} T.~P.,  {Stern} D.,  {Tao} L.,    {Webb} N.,
  2016a, \apjl, 827, L13

\bibitem[\protect\citeauthoryear{{Walton}, {F{\"u}rst}, {Bachetti}, {Barret},
  {Brightman}, {Fabian}, {Gehrels}, {Harrison}, {Heida}, {Middleton}, {Rana},
  {Roberts}, {Stern}, {Tao} \& {Webb}}{{Walton} et~al.}{2016b}]{walton17}
{Walton} D.~J.,  {F{\"u}rst} F.,  {Bachetti} M.,  {Barret} D.,  {Brightman} M.,
   {Fabian} A.~C.,  {Gehrels} N.,  {Harrison} F.~A.,  {Heida} M.,  {Middleton}
  M.~J.,  {Rana} V.,  {Roberts} T.~P.,  {Stern} D.,  {Tao} L.,    {Webb} N.,
  2016b, \apjl, 827, L13

\bibitem[\protect\citeauthoryear{{Walton}, {F{\"u}rst}, {Harrison}, {Stern},
  {Bachetti}, {Barret}, {Brightman}, {Fabian}, {Middleton}, {Ptak} \&
  {Tao}}{{Walton} et~al.}{2018}]{walton18}
{Walton} D.~J.,  {F{\"u}rst} F.,  {Harrison} F.~A.,  {Stern} D.,  {Bachetti}
  M.,  {Barret} D.,  {Brightman} M.,  {Fabian} A.~C.,  {Middleton} M.~J.,
  {Ptak} A.,    {Tao} L.,  2018, \mnras, 473, 4360

\bibitem[\protect\citeauthoryear{{Walton}, {F{\"u}rst}, {Heida}, {Harrison},
  {Barret}, {Stern}, {Bachetti}, {Brightman}, {Fabian} \& {Middleton}}{{Walton}
  et~al.}{2018}]{walton18b}
{Walton} D.~J.,  {F{\"u}rst} F.,  {Heida} M.,  {Harrison} F.~A.,  {Barret} D.,
  {Stern} D.,  {Bachetti} M.,  {Brightman} M.,  {Fabian} A.~C.,    {Middleton}
  M.~J.,  2018, \apj, 856, 128

\bibitem[\protect\citeauthoryear{{Walton}, {Pinto}, {Nowak}, {Bachetti} \& {et
  al.}}{{Walton} et~al.}{2020}]{walton20}
{Walton} D.~J.,  {Pinto} C.,  {Nowak} M.,  {Bachetti} M.,    {et al.} 2020,
  \mnras, 494, 6012

\bibitem[\protect\citeauthoryear{{Wang}, {Soria} \& {Wang}}{{Wang}
  et~al.}{2019}]{wang19}
{Wang} C.,  {Soria} R.,    {Wang} J.,  2019, \apj, 883, 44

\bibitem[\protect\citeauthoryear{{Weng} \& {Feng}}{{Weng} \&
  {Feng}}{2018}]{weng18}
{Weng} S.-S.,  {Feng} H.,  2018, \apj, 853, 115

\bibitem[\protect\citeauthoryear{{Zdziarski}, {Johnson} \&
  {Magdziarz}}{{Zdziarski} et~al.}{1996}]{zdiarski96}
{Zdziarski} A.~A.,  {Johnson} W.~N.,    {Magdziarz} P.,  1996, \mnras, 283, 193

\end{thebibliography}

\bsp
\label{lastpage}
\end{document}